\documentclass{thesis}
\usepackage{graphicx}% if you want to include graphics files
\usepackage{import}
\usepackage{listings}
\usepackage{seqsplit}
\usepackage{algorithm}
\usepackage{algpseudocode}
\usepackage{url}
\usepackage{graphicx}
\usepackage[utf8]{inputenc}
\usepackage{amsmath}% http://ctan.org/pkg/amsmath
\usepackage{cite}
\usepackage{subcaption}

\thesistitle{\bf A survey of data transfer and storage techniques in prevalent cryptocurrencies and suggested improvements}        
\author{Sunny Katkuri}        
\degree{Master of Science}
\department{Computer Science} % provide your area of study here; e.g.,
\projadviser{Prof. Brian Levine}
\submitdate{May 2018\\}        
\begin{document} 
\titlepage             % Print titlepage   
%\copyrightpage        % optional         
\tableofcontents       % required 
\listoftables          % required if there are tables
\listoffigures         % required if there are figures

\specialhead{ACKNOWLEDGMENT}

Prof. Brian Levine's course and his holistic approach  started my fascination towards the different domains that blockchains and cryptocurrencies encompass, both technical and social. Working under him since then has been enriching and his guidance and experience give me much-needed focus in my meandering ways.
\\
\\
This work would not have been possible without the countless open-source code contributors silently chipping away at projects they believe in far away from the deafening crowds and hype. \\
\\
Thank you and keep bitcoin weird.

\specialhead{ABSTRACT}
This thesis focuses on  aspects related to the functioning of the gossip networks underlying three relatively popular cryptocurrencies: Ethereum, Nano and IOTA. We look at topics such as automatic discovery of peers when a new node joins the network, bandwidth usage of a node, message passing protocols and storage schemas and optimizations for the shared ledger. We believe this is a topic that is often overlooked in works about blockchains and cryptocurrencies.  Vulnerabilities and inefficiencies attain a higher significance than ones in a regular open source project because of the rather direct financial implications of these projects. Barring Bitcoin, a network that has been around for nearly 10 years,  no other project has substantial documentation for its operational details other than scattered and sparse pages in the source code repositories. Almost all of the content described here has been extracted by studying the source code of the reference implementations of these projects. 

We evaluate the use of Invertible Bloom Lookup Tables and the Graphene protocol to decrease block propagation times and bandwidth usage of certain messages. We perform realistic simulations that show significant improvements. We provide a complete implementation of Graphene in Geth, Ethereum's main node software and test this implementation against the main Ethereum blockchain.  

We also crawled the chosen cryptocurrency networks for publicly visible nodes and provide an Autonomous System-level breakdown of these nodes with the end goal of estimating the ease of performing attacks such as BGP hijacks and their impact.

Code written for implementing Graphene in Geth, performing various simulations and for other miscellaneous tasks has been uploaded to Github at \url{https://github.com/sunfinite/masters-thesis}.

\chapter{INTRODUCTION}
\section{Background}
Consensus in distributed systems has been well-studied for at least four decades. Seminal results such as the Byzantine Generals Problem\cite{byzantine} and the FLP impossibility\cite{flp} date back to 1982 and 1985 respectively. The authors of the FLP result formalize consensus to achieving the following three properties in a system 
\cite{papertrailflp}:
\begin{itemize}
    \item \textbf{termination}: all participants decide on a state eventually;
    \item \textbf{agreement}: all participants decide on the same value for the state;
    \item \textbf{validity}: the value must have been proposed by one of the participants and not arrived at by default.
\end{itemize}

Consensus in the presence of misbehaving or malicious participants was the crux of this early research. The Byzantine Generals Problem states this scenario as a group of military generals who have to decide on a single time to launch an attack. They do this by passing messages and there might be traitors who willingly distort not just their own messages but also other messages passing through them. The FLP impossibility states that consensus cannot be achieved in an asynchronous system where any system can fail silently. The solution to the Byzantine Generals Problem assumes synchronicity and puts a limit of $2k + 1$ honest generals in the presence of $k$ traitors to achieve consensus. Later consensus protocols such as Paxos\cite{paxos} and Raft\cite{raft} also require leader election and knowledge of the participants to achieve consensus. The problem with these systems is that they do not perform efficiently at the scale of an open network like the Internet.

Bitcoin\cite{bitcoin-whitepaper} sidestepped the requirement of knowing other participants beforehand by having participants follow the general who presents a cryptographic proof of the most power in a round of consensus. As long as a simple majority of participants stay honest and follow the right winner, the state of the system achieves consensus.

To become the winner of a consensus round, a participant has to provide the solution to a hard cryptographic puzzle. The chances of finding a solution are only dependent on the amount of compute power the participant puts into the problem. Upon winning a round, the participant gets to decide the next update to the shared state that is being  maintained locally in all participants. Each update consists of set of transactions transferring value from one public key to another. These transactions grouped together form a block. And since these are updates to a previous balance of a public key, each block has to refer to a previous block that it is updating in its proof, thus forming a chain of blocks i.e.\ a blockchain. The winner of a round gets a pre-determined amount of newly minted coins --- they have now \textit{mined} new coins into existence --- as a reward for putting in the work behind producing the block.

Since any participant can work towards producing the next block, they need to have knowledge of all ongoing transactions that can possibly be included in the next block. Blockchains achieve this scenario by implementing a peer-to-peer gossip network where every transaction propagates to every known node on the network. 

\subsection{Graphene}
Graphene \cite{graphene} is a protocol to reduce the bandwidth consumption during block propagation in a peer-to-peer blockchain. 
Currently, when a new block is propagated to a peer, all the transactions in that block are re-transmitted even though the gossip network would have already broadcast these transactions with a high probability when they were originally submitted. 

Two prior solutions have been proposed to avoid this duplicate data transfer in Bitcoin: Compact Blocks\cite{compact} and X-treme Thin Blocks\cite{xtreme}. Instead of including full transactions, Compact Blocks only send 6-byte short transaction IDs. The peer then makes a special request for transactions not present locally. In X-treme Thin Blocks, the receiver first builds a Bloom filter from the transactions present with them and transmits it to the sender. The sender checks this Bloom filter to determine which transactions need to be sent in full. IDs of all transactions in the block are also sent.

Graphene reduces the cost of propagation even further than the above two protocols by never sending the full list of transaction IDs. Instead, a special data structure called Invertible Bloom Lookup Table (IBLT)\cite{iblt} is constructed by both parties wanting to exchange a block. The peer with the block transmits this encoded data structure which is then used by the other peer to perform set reconciliation with its own local IBLT. The goal of efficient set reconciliation is to find the union $S_i \cup S_j$ of two sets $S_i$ and $S_j$ whose symmetric difference $d = \mid S_i - S_j \mid + \mid S_j - S_i \mid$ is low when compared to the cardinality of $S_i$ and $S_j$. IBLTs require only $O(d)$ space to find the difference instead of $O(n)$. Graphene also uses a Bloom filter to reduce the number of transactions added to the local IBLT at the receiver. This improves the chances of recovering the difference from the remote IBLT. For a listing of the steps followed by Graphene for propagating blocks and a detailed description of Bloom filters and IBLT, see Appendix \ref{appendix:graphene}.

Transactions that are not yet included in a block are stored in the peer's $mempool$. In a well-connected network, these $mempool$s should be in sync with a very high probability. Graphene relies on this property to formulate block propagation as a set reconciliation problem. The cost of propagation a block in graphene in given by:

$$T(a) = n\frac{-ln(\frac{a}{m-n})}{8ln^2(2)} + ad\tau$$

where,  $m$ = number of transactions in the receiver's $mempool$,\\
$n$ = number of transactions in the block, \\
$a$ = expected symmetric difference between the two IBLTs, \\
$\tau$ = cost in number of bytes of each cell of an IBLT, \\
and $d$ = Constant factor introduced to increase in the number of cells in the IBLT to improve chances of recovery\cite{iblt-constant}.\\

In the equation above, the left operand signifies the cost of the Bloom filter and the right operand signifies the cost of the IBLT. 

\section{Contributions}

\subsection{Documentation and Improvements}
We provide comprehensive documentation of the protocols behind various networking and operational aspects of three cryptocurrencies: Nano, Ethereum and IOTA. We know of no similar undertaking for these projects currently. We also propose and evaluate improvements to different parts of these protocols. The improvements along with a brief summary of the cryptocurrencies are described below.

\subsubsection{Nano}
Nano\cite{nano} tries to remove the latency inherent in Proof of Work blockchains where transactions are confirmed only when a block is produced by a miner. It uses a model similar to Delegated Proof of Stake(DPoS)\cite{dpos} where each account nominates a representative to vote on its behalf. Voting takes places to resolve conflicts about updates to the shared ledger. Votes are included as a part of the updates themselves rather than as a separate process. Nano also breaks down a single block of transactions into small blocks containing just one transaction. In effect, this amounts to each account having its own blockchain. This structure reduces the contention for shared resources provided by the participants in the network. Unlike Bitcoin where a small reward is provided to the miner who provides a valid block Nano also does not have incentives for confirming transactions. Active users themselves have to participate in the upkeep of the network thereby reiterating the $Stakeholder$ model. The rationale behind this is that once a participant owns a high enough stake of the value in the network, they are automatically vested in ensuring its proper functioning since any aberration would mean a decrease in the value of their stake.

In Section~\ref{sec:nano-keepalive} we list a vulnerability we discovered in Nano's node discovery protocol. We were able to eclipse all connections made by Nano node and also make it unresponsive in under 10 minutes by just executing one instance of the attack requests. Given Nano's small network footprint (detailed in Section~\ref{sec:nano-network-analysis}), the entire network can be attacked to induce massive delays using just a few machines. This vulnerability has been reported to the development team and a patch is being implemented at the time of writing.

In Section~\ref{sec:nano-frontier} we evaluate a proposal to use IBLTs in a key periodic sync message sent by Nano. Nano's bandwidth usage has already been the subject of complaints \cite{nano-bandwidth-1} \cite{nano-bandwidth-2}. We show that using IBLTs produces a decrease of three orders of magnitude in bytes transferred and also speeds up the processing of this message by about the same amount.

\subsubsection{Ethereum}

Ethereum\cite{ethereum} is the second-most valuable cryptocurrency after Bitcoin. It currently follows the same underlying principles of operation as Bitcoin wherein nodes have to perform computation to produce a block. The key differentiating factor for Ethereum is its ability to treat transactions as full-fledged computer programs rather than just denoting a transfer of coin. This lets us run arbitrary code whose execution and data can be publicly verified on the blockchain. These programs called Smart Contracts have been used to implement a wide range  of applications from decentralized mutual funds to collectible cats which are guaranteed to be digitally unique.

In Section~\ref{sec:eth-experiment}, we detail the implementation and evaluation of using Graphene to propagate blocks in Geth, Ethereum's reference node implementation written in Golang. Over 60\% of public Ethereum nodes run Geth. Ethereum has a much faster block production time than Bitcoin ($\approx 15\ seconds$) and has no implemented feature to optimize block propagation (like Compact Blocks in Bitcoin). Ethereum also has a higher transaction rate because of automated calls from Smart Contract executions. This increases Graphene's relevance in reducing latencies and bandwidth usage on the Ethereum network to keep up with the block and transaction rate.

\subsubsection{IOTA}

Similar to Nano, IOTA does not rely on the standard Bitcoin view of the blockchain but instead uses a Directed Acyclic Graph (DAG) of transactions to achieve consensus. This DAG is called the `Tangle' and relies on each transaction referring to two previous transactions in the Tangle as a way of providing confirmation to these transactions. A transaction's confidence score is directly tied to the number of transactions which reference it and is similar to the concept of block depth in Bitcoin. A new transaction chooses an old transaction to refer by performing a weighted random walk starting from the genesis transaction. Section~\ref{sec:iota-sync} provides more details about this process.

In Section~\ref{sec:iota-experiment}, we evaluate the use of cuckoo filters to implement a fast lookup cache to check if a transaction can be used as a part of a snapshot to update account balances. This eliminates two disk reads per transaction.

\subsection{Network Analysis}

Attack vectors like BGP hijacking\cite{bgp-hijacking} on public cryptocurrency networks can have a significant impact if they succeed in cutting off a considerable fraction of online nodes. Such attacks could also feed false messages and views of the blockchain to the targeted section of nodes. There have been only a couple of studies looking at centralization in Bitcoin at the Autonomous System (AS) level. Feld et al.~\cite{bitcoinas} pointed out that only 10 ASes contain as many 30\% of all public Bitcoin nodes. Apostolaki et al.~\cite{hijack} described the design of partitioning and delay attacks that could be performed using the topology that existed at that time. They found found that nodes which account for almost 50\% of the blocks produced in Bitcoin could be isolated by hijacking only 39 prefixes. They were even able to demonstrate diversion of traffic meant for their own AS via a false BGP announcement in under 2 minutes. Analysing BGP adversitements, they found that at least 100 Bitcoin
nodes are victims of hijacks each month.
Other attempts to subvert the Bitcoin network via false BGP advertisements have already been observed\cite{bitcoinbgpattack}. In this attack, the attackers were able to successfully steal the mining work performed by a portion of a pool.

We perform a similar AS-level analysis for the three cryptocurrencies we are considering. We also provide a country-level breakdown of nodes because the number of participants in a cryptocurrency without proof-of-work incentives might serve as a reliable estimate of the adoption of the currency for purposes other than mere speculation.

\subsection{Transaction ordering in Graphene}

Graphene can reliably determine which transactions from a node's mempool belong to a block. It can also recover transaction IDs which may not be present in our mempool. But it does not provide information about how the transactions are ordered within a block. This is a consequence of the randomized hashing performed during insertion into the IBLT. Ordering is critical if more than one transaction updates the same account in the same block. Ordering is also needed to generate the right Merkle root that is included in the block header.  Graphene proposes an $O(n)$ solution that relies on a pre-determined sort function for transaction IDs (eg. lexicographicallly) and then transmitting $nlog(n)$ bits containing the indices. We propose and evaluate an alternate idea that seeks to avoid sorting and also utilize IBLT's \textit{valueSum} field and the same hashing procedure to encode index positions. We transaction indices are encoded in this fashion, procedures to solve Constraint Satisfaction Problems (See chapter 5 of \cite{aima}) can be used to decode their values. 

Our solution utilizes the \textit{valueSum} field of the graphene IBLT. Each \textit{valueSum} field is further divided into $b$ buckets where each bucket is large enough to hold a transaction index. This gives us the value of $b$ as:
	$$b = \frac{v}{\frac{\lceil log_2 n\rceil}{8}}$$
	where $v$ is the number of bytes per \textit{valueSum} field.

When a transaction is inserted into the IBLT, it is added to $k$ cells. For each transaction t  inserted into cell $i$, its index bucket $b_{ti}$ is given by:
$$b_{ti} = t_i \bmod b$$
where $t_i$ is the $i^{th}$ byte of transaction $t$.

At the senders end, the value of the index bucket is updated to:
$$b_{ti} = b_{ti} + I(t)$$
where $I(t)$ is the index of transaction $t$.

Assuming that the $k$ hash functions are independent and the transaction IDs are generated using a uniformly distributed hash function, each bucket will be the sum of $n / bk$ transaction indices.

At the receivers end, $b_{ti}$ is the RHS of a linear constraint. The receiver generates a constraint by keeping track of all transactions $t_{1..p}$ that fall into an index bucket: 
$$I(t_1) + I(t_2) + \ldots + I(t_p) = b_{ti}$$

We now define the parameters of the CSP:\\
\textbf{Variables}:$ I(t_1), \ldots, I(t_n)$ \\
\textbf{Domain}: 1 to n \\
\textbf{Constraints}: $l$ linear equations where $l$ lies between $k + 2$  and $kb + 2$. The two constant constraints are:
$$\sum_{i=1}^{n} I(t_i) = n (n + 1) / 2$$
$$MerkleRoot(t_1, \ldots, t_n)= merkle\_root\_in\_block\_header$$

\subsubsection{Evaluation}
\begin{figure}
\caption{Successful index recoveries without guessing}
\label{img:csp_no_guess}
\includegraphics[scale=0.7]{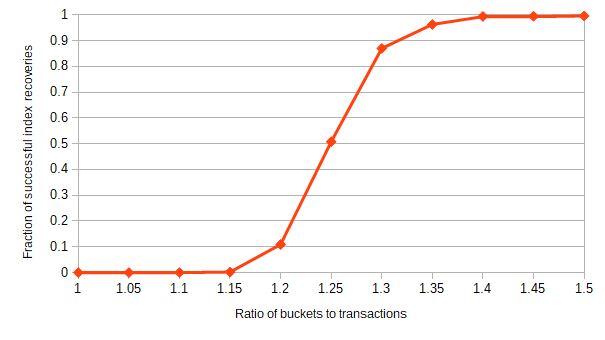}
\centering
\end{figure}

We simulated addition of transaction indices to IBLTs by generating random transaction IDs, ordering them and  encode the order as index sums in the IBLT. At the other end, the procedure was repeated to find the IDs which map to a specific bucket. Once the equations have been generated, we remove the unary constraints (equations with only one variable) and substitute its value in the other equations. We then recursively simplify equations using the same procedure. Decoding all index values in this fashion is our ideal case since we did have not guess any value. Figure~\ref{img:csp_no_guess} shows that this happens with a very high probability when the number of buckets is at least 1.3 times the number of transactions. 1000 trials with different transaction counts were performed for each bucket to transaction count ratio.

To bring this ratio down to 1, we considered two ideas. The first idea is to reduce this to a problem of finding $n$ unknowns using $n$ equations and then solving the matrix $Ax = b$. To do this, we first recursively simplify all unary constraints. We then send unencoded transaction indices of the minimum number of transactions required to reduce the number of unknowns to be equal to the number of equations. Table~\ref{tbl:unencoded-indices} lists the number of unencoded indices that had to be sent for different bucket ratios with the number of transactions fixed at 400 (which is a comfortable upper bound for Ethereum blocks, see Section~\ref{sec:eth-experiment})

\begin{table}[]
\centering
\caption{Number of unencoded indices transferred for 400 transactions}
\label{tbl:unencoded-indices}
\begin{tabular}{|lllll|}
\hline
\textbf{1.0}   & \textbf{0.95}     & \textbf{0.9}  & \textbf{0.85} & \textbf{0.8} \\
\hline
25   & 36 & 57 & 71 & 85\\
\end{tabular}
\end{table}

The second approach we considered is to use depth first search backtracking with forward checking and minimum remaining value (MRV) heuristics. This technique is used to solve standard Constraint Satisfaction Problems such as map coloring. Using our bare bones simulation, ee were not able to generate the correct assignment within a reasonable amount of guesses and the additional computation required to continue the inference process negates any processing or bandwidth gains over sorting or other methods.

\subsection{Evaluating Graphene for unsynchronized mempools}

Graphene assumes a very high similarity between mempool of two nodes exchanging blocks. We relaxed this assumption and calculated the bytes required by averaging data from 5 randomly chosen real Bitcoin blocks. This scenario may occur if a node has suffered a network outage for an extended period of time or has a very high round trip times.

\begin{figure}
\caption{Bytes transferred during block propagation}
\label{img:mempool}
\includegraphics[scale=0.7]{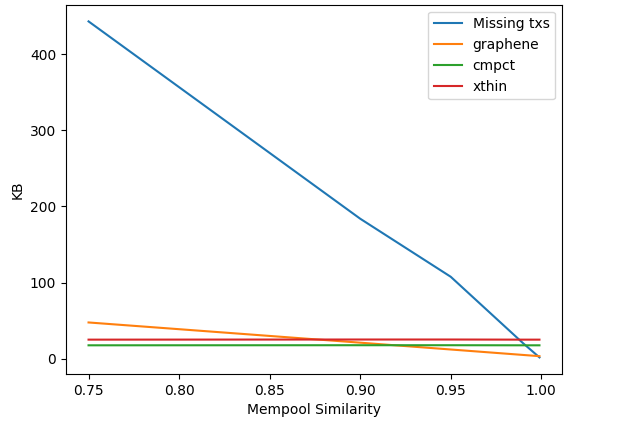}
\centering
\end{figure}

\begin{figure}
\caption{Number of missing transactions}
\label{img:n_missing}
\includegraphics[scale=0.7]{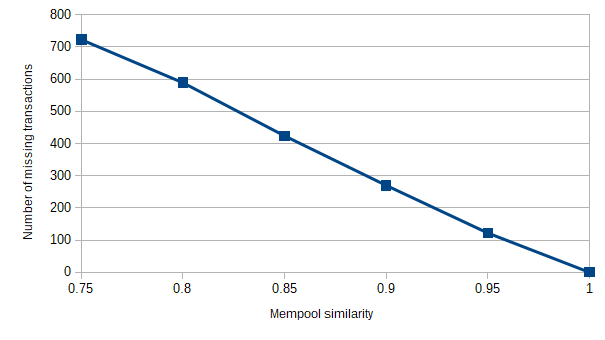}
\centering
\end{figure}

Figure\ref{img:mempool} shows the number of bytes required by the three efficient block propagation protocols that we have seen. It shows that Graphene fares worse than Compact and Xthin blocks when the mempool differ by over 10\% though the cost of transferring the encoded block is eclipsed by the number of bytes required to fetch the actual missing transactions as the similarity decreases. Figure\ref{img:n_missing} shows the the number of missing transactions versus mempool similarity averaged over the same 5 blocks.

\chapter{Nano}
\section{Node Discovery}
A client running Nano starts by connecting to pre-configured bootnodes obtained by resolving \texttt{rai.raiblocks.net}. $keepalive$ messages are sent over UDP to each of the resolved IP address. Only IPv6 addresses are considered and any IPv4 address is mapped to its IPv6 equivalent. Peers respond with another $keepalive$ message on receipt of a $keepalive$ message. Each $keepalive$ message includes the addresses of at most eight peers chosen randomly from the sender's peer list. The receiver checks if each address in the received list is a valid peer address and if it is a new peer, a $keepalive$ message is sent and the process repeats. There is no restriction placed currently on the number of peers a client can have. The protocol version being used is included in the header of all messages transmitted in Nano. This version is validated against the minimum protocol version supported at the receiver and the peer is dropped otherwise. The keepalive procedure is repeated every 60 seconds. For each run of the procedure, peers that haven't been heard from in the last 300 seconds are dropped . A $keepalive$ message is also sent when we receive a $publish$, $confirm\_req$ or $confirm\_ack$ message.

\section{Block Synchronization}
\label{sec:nano-sync}

Before we delve into how blocks are synchronized, let's take a look at the different types of blocks in Nano. A block is defined to contain only one transaction such that it can fit into a single UDP packet (65535 bytes). In contrast to other blockchains, blocks are tied to a specific account and are signed by the private key corresponding to the account ID, which is the base-58 encoding of a public key. 

\textbf{Send block}: A $send$ block is created when value is being transferred to someone. This block has to refer a previous block belonging to the account that initiated the transfer. The other fields are the destination account ID and the amount being transferred.

\textbf{Open block}: If the destination account in a $send$ block does not exist on the blockchain, an $open$ block has to be published to claim the amount. An $open$ block does not have a reference to a previous block but instead refers to the send block as the $source$. The other fields are the ID of the account being created and representative account who will vote on the new account's behalf.  An account can be its own representative.

Note that for a new account to appear on the blockchain it has to have a valid endorsement from an existing account in the form of a $send$ block. The genesis block is a special $open$ block (Listing~\ref{lst:nanogenesis}).

\begin{lstlisting}[caption=Nano genesis block, basicstyle=\small, escapechar=\%, label={lst:nanogenesis}]
{
    type: open,
    source: %{\scriptsize\seqsplit{E89208DD038FBB269987689621D52292AE9C35941A7484756ECCED92A65093BA}}%,
    representative: %{\scriptsize\seqsplit{xrb\_3t6k35gi95xu6tergt6p69ck76ogmitsa8mnijtpxm9fkcm736xtoncuohr3}}%,
    account: %{\scriptsize\seqsplit{xrb\_3t6k35gi95xu6tergt6p69ck76ogmitsa8mnijtpxm9fkcm736xtoncuohr3}}%,
    work: 62f05417dd3fb691,
    signature: %{\scriptsize\seqsplit{9F0C933C8ADE004D808EA1985FA746A7E95BA2A38F867640F53EC8F180BDFE9E2C1268DEAD7C2664F356E37ABA362BC58E46DBA03E523A7B5A19E4B6EB12BB02}}%
}
\end{lstlisting}

The source block being referenced does not exist on the blockchain. The balance for the genesis account is set to $2^{128} - 1$, which is the total number of Nano that can ever be in circulation.

The message included in the signature is the blake2b hash of the concatenation of account ID, representative ID and the source block hash.

\textbf{Receive block}:  A $receive$ block is created by the destination account of a pending $send$ block. On a successful publish of this block, the corresponding $send$ block is said to be ‘pocketed’. The fields in this block include the hash of the $send$ block as the source and the hash of the previous block belonging to the destination account.

\textbf{Change block}: A $change$ block is published when an account wants to change the delegated representative. The fields in this block include the new representative account ID and the hash of the previous head block of the account being modified.

\textbf{State block}: A new type of block which seeks to unify the different block types into a single container. The fields in this block include the ID of the account publishing this block, the balance for this account, the hash of the previous block belonging to this account, representative ID and a $link$ field which refers to the source block hash in a receive scenario or to a destination account in a send scenario.

Since nano does not have a canonical block height, the method chosen to enable new non-backwards-compatible features is to generate pre-determined transactions and have the code check for the presence of these transactions. State blocks are being enabled via two such $canary$ blocks: one to let nodes that state blocks can be parsed and another to let nodes generate state blocks. At the time of writing, only the parse canary block has been published.

\subsection{Bootstrapping}

On node boot, a certain number of peers with network version greater than 5 are chosen and contacted for synchronization. The target number of peers to contact is determined using the method listed in Algorithm ~\ref{contact-nano}.

\begin{algorithm}
    \caption{Number of peers to contact on node start}
    \label{contact-nano}
\begin{algorithmic}[1]
\State $minPeers \gets 4 $ \Comment Can be changed at start
\State $maxPeers \gets 64$ \Comment Can be changed at start
\State $estBlocksPerBootstrap \gets 50000$
\Statex
\Procedure{GetBootstrapPeerCount}{$pullsInProgress$}

\State $step \gets pullsInProgress / estBlocksPerBootstrap$
\State $target \gets minPeers + (maxPeers - minPeers) * step$
\State \Return \Call{max}{$1,target$}
\EndProcedure
\end{algorithmic}
\end{algorithm}
    
If the node is still waiting for peers, this selection is attempted every 5 seconds for the first 3 times (aggressive warmup) and cools down to every 300 seconds subsequently. The actual number of new bootstrap connections to be established is further determined using the function listed in Algorithm~\ref{alg:actualConns}.

\begin{algorithm}
    \caption{Number of bootstrap connections actually established}\label{alg:actualConns}
\begin{algorithmic}[1]
\State $target \gets $ \Call{GetBootstrapPeerCount}{$pullsInProgress$} 
\Statex
\Procedure{GetNewConnectionCount}{$activeConnCount$} 
\Statex \Comment $activeConnCount$ includes both $pullsInProgress$ and idle connections not yet scheduled for a pull
\State $maxNewBoostrapAttempts \gets 10$
\State $connCount = (target - activeConns) * 2$
\State \Return \Call{min}{$connCount,maxNewBootstrapAttempts$}
\EndProcedure
\end{algorithmic}
\end{algorithm}

The target number of peers obtained above is doubled with the rationale that not many peers respond to bootstrap requests and hence more attempts than necessary have to be made. There is no check whether a peer has already been contacted when selecting and even though the time of the last bootstrap attempt is maintained per peer it is currently not utilized anywhere in code. Once a connection is established, it is added to an idle queue.

The bootstrap process then picks a connection from the idle queue and makes a frontier request over it.  The peer responds with all account IDs in its blockchain and the corresponding block hash of the latest block for that account (the head block). The request may optionally include an age for accounts and count but these are set to $INT\_MAX$ for bootstrap requests. Each ID, hash pair results in a socket write/read operation. The node keeps track of the read rate and if it falls below 1000 pairs per second the frontier request is aborted. If this is not the first time that the node is being started, there may be accounts stored locally. The returned account is compared with the latest local account and if the local account is greater, it means that the bootstrap peer does not know about this account --- this is because the database stores accounts in a lexicographically sorted order --- and head blocks of accounts greater than the returned account are added to an unsynced list which is used for a bulk pushed later. If the returned and local accounts match and the returned block hash is already in the local blockchain, then a bulk push is scheduled for the missing block hashes. We haven't actually pulled the real blocks themselves, so if we have gotten this far and a push wasn't needed, a pull is scheduled for this account with the block hash returned in the frontier response. 

After the frontier request completes, the idle connection pool is now utilized to run the scheduled bulk block pulls. The connection pool does not remain static. If a connection’s block fetch rate falls below 10 blocks per second ($\approx1.5Kbps$), it is removed from the connection pool and the pool is repopulated using the target calculated above. While repopulating, if the current pull connections count is more than $2/3$ of the new target then the slowest $\sqrt{target}$ connections and dropped and new connections established. This metric has been picked arbitrarily and could use some tuning.

After all the pulls are complete, the unsynced list is iterated and each block and its dependencies (previous or source blocks) are pushed to the peer. The same connection pooling strategy is used as before. In the block propagation section, we will take a look at how a block is validated and confirmed.

All bootstrap related messages and data are transmitted via TCP. Other node to node communication is via UDP over the same port.

\section{Block Propagation}
For a block to validated, the hash of its root and work value (a 64-bit integer) should be less than publish threshold. This threshold is currently \texttt{0xffffffc000000000} for the main network which is estimated to require at least 5 seconds of calculations on a reasonably powerful laptop. This acts as the rate limit for blocks and hence transactions.
 
As soon as a valid block is observed, an election for this block is started by having all representative accounts local to the node’s wallet vote for this block. The block is rebroadcast using a plain publish message if there are no local representatives. The number of peers for rebroadcast is set to the square root of the total number. If there are local representatives, a $confirm\_ack$ message is sent containing the vote for this block. This announcement cycle is repeated every 16 milliseconds and the number of announcements across all elections are limited to 32 per cycle as a flow control mechanism. After four successful announcements have been made for a block, its vote is tallied and if the weights of representatives that have voted for it exceeds quorum, the block is considered confirmed. A representative's weight is the sum of its own balance and the balances of all other accounts that have designated it as a representative.  Quorum is currently set at more than half of total valid supply. The remaining balance of the genesis account and coins sent to the burn address (all 0s) are excluded from the valid supply.
 
A fork is detected when the current block’s previous block is not the frontier transaction for this account. Upon detection, $confirm\_req$ messages for both the current block and the previous blocks immediate confirmed successor are broadcast to selected peers (see representative crawling below). Upon receiving a $confirm\_req$ message, the peer will start an election of its own for this block and respond with a $confirm\_ack$ message for any existing successor of the current block's previous block.

\subsection{Representative crawling}
 
$confirm\_req$ messages are broadcast only to those peers which might have representative accounts and can vote in an election. This is determined by periodically sending $confirm\_req$ messages for a randomly chosen block to all peers and checking the responses for a $confirm\_ack$ message. The peer’s possible representative account and its weight are stored.

\section {Block Storage}

Nano uses Lightning DB\cite{lmdb}, a memory-mapped persistent key-value store. Memory mapping limited databases to only 4GB on 32-bit architectures. This limitation does not exist on 64-bit architectures and Nano uses a memory mapped file that can grow up to 1TB.

Table~\ref{nano-schema} lists the different types of keys that can occur and the format of their values.

\begin{table}[]
\centering
\caption{Schema of tables in Nano}
\label{nano-schema}
\begin{tabular}{|l|l|p{7cm}|}
\hline
\multicolumn{1}{|c|}{\bfseries Name} & \multicolumn{1}{c|}{\bfseries Key} & \multicolumn{1}{c|}{\bfseries Value} 
\\ \hline
frontiers      & block\_hash & account                                                                         \\ \hline
account        & account     & \textless{}head\_block hash, representative, balance, last change\textgreater{} \\ \hline
send\_blocks   & block\_hash & send\_block                                                                     \\ \hline
receive\_block & block\_hash & receive\_block                                                                  \\ \hline
open\_blocks   & block\_hash & open\_block                                                                     \\ \hline
change\_block  & block\_hash & change\_block                                                                   \\ \hline
pending        & block\_hash & \textless{}sender, amount, destination\textgreater{}                            \\ \hline
blocks\_info   & block\_hash & \textless{}account, balance\textgreater{}                                       \\ \hline
representation & account     & weight                                                                          \\ \hline
unchecked      & block\_hash & block                                                                           \\ \hline
unsynced       & block\_hash &                                                                                 \\ \hline
vote           & account     & uint64                                                                          \\ \hline
\end{tabular}
\end{table}

\section{Network Analysis}
\label{sec:nano-network-analysis}

We crawled the Nano network by sending only $keepalive$ messages as these already include a list of at most eight peers. We were able to detect over 500 public nodes. This data was validated and augmented by other public sources \cite{nanopeers}. ASN and geoip data was fetched from RIPE\cite{ripe}, Routeviews\cite{routeviews} and Geolite\cite{geolite}

\begin{figure}
\caption{Country-wise distribution of 541 public Nano nodes}
\label{img:nano-country}
\includegraphics[scale=0.6]{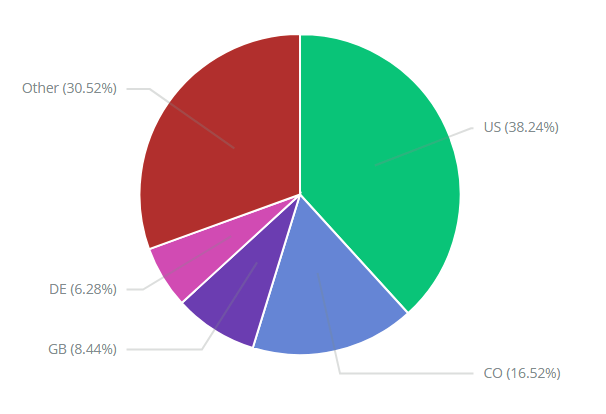}
\centering
\end{figure}

Figure~\ref{img:nano-country} shows the distribution of nodes by country. Colombia is a surprising presence on the list.

\begin{figure}
\caption{ASNs containing public Nano nodes}
\begin{subfigure}[b]{0.5\textwidth}
\caption{CDF}
\label{img:nano_asn_cdf}
\includegraphics[width=\textwidth]{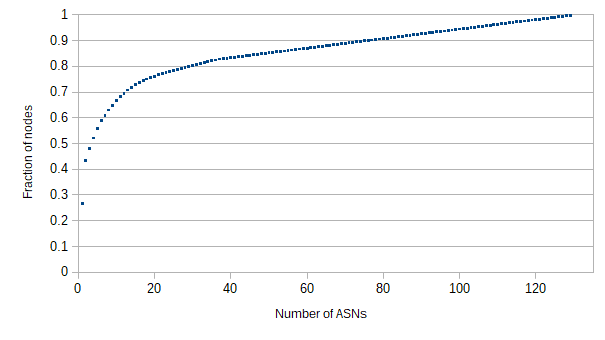}
\end{subfigure}
\begin{subfigure}[b]{0.5\textwidth}
\caption{Count}
\label{img:nano_asn_hist}
\includegraphics[width=\textwidth]{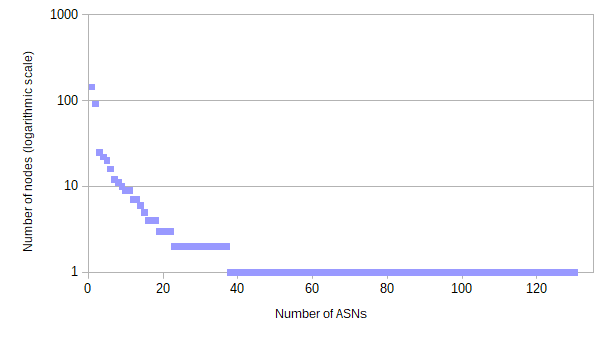}
\end{subfigure}
\end{figure}

ASN 14061 contains over a quarter of Nano's nodes(Table ~\ref{tbl:nano-asn-table}). Nano's node distribution has a very long tail(Figure~\ref{img:nano_asn_hist}).
\begin{table}[]
\centering
\caption{Top 5 ASNs with public Nano nodes}
\label{tbl:nano-asn-table}
\begin{tabular}{|lp{7cm}l|}
\hline
\textbf{ASNum}   & \textbf{Name}     & \textbf{Percentage  of nodes} \\
\hline
14061 & DIGITALOCEAN-ASN - DigitalOcean & 26.6 \\
14080 & Telmex Colombia S.A. & 16.8 \\
16276 & OVH - OVH SAS & 4.6 \\
24940 & HETZNER-AS - Hetzner Online GmbH & 4 \\
16509 & AMAZON-02 - Amazon.com & 3.7\\
\hline
\end{tabular}
\end{table}

\section{Networking Protocol}
This section lists the various messages transmitted by nodes running Nano. We've already covered these messages while describing the node operation above. The following serves as a formal listing.

A header is included in every message that is sent. The header fields are as follows:
\begin{enumerate}
    \item \texttt{magic\_number}: The letters $RC$ for the main network
    \item \texttt{version\_max}: Maximum version of the messaging and consensus protocol supported by this client
    \item \texttt{version\_min}
    \item \texttt{version\_using}
    \item \texttt{type}: An integer indicating which of the messaged described below is being sent.
    \item \texttt{extensions}: A single byte to indicate any special handling for this message.
\end{enumerate}	

\subsection{Messages}
\begin{enumerate}
    \item \textbf{keepalive} \\
    This message is sent to inform a peer of our presence and our knowledge of other peers. \\
    \textit{Fields}:
        \begin{enumerate}
            \item \texttt{peers}: A list of up to 8 peers.
        \end{enumerate}
        
    \item \textbf{publish} \\
    This message is sent when a new transaction occurs. Each transaction is wrapped in its own block. \\
    \textit{Fields}:
        \begin{enumerate}
            \item \texttt{block}: Please refer to Section~\ref{sec:nano-sync} for a description of the different fields present in a block.
        \end{enumerate}
    
    \item \textbf{frontier\_req} \\
    This message is sent to a peer to fetch the latest block (frontier) for an account or a list of accounts. \\
    \textit{Fields}:
        \begin{enumerate}
            \item \texttt{start}: The request is for frontiers of accounts greater than or equal to this field. As mentioned before, accounts are stored on disk in a lexicographic manner.
            \item \texttt{age}: The response should only contain frontiers for accounts which have changed in the last \texttt{age} seconds.
            \item \texttt{count}: Maximum number of frontiers to be sent in the response.
        \end{enumerate}
    
    \item \textbf{bulk\_pull} \\
    A $bulk\_pull$ message is sent for each account that is determined to be out of sync via $frontier\_req$. \\
    \textit{Fields}:
        \begin{enumerate}
            \item \texttt{account}
            \item \texttt{end}: The response should contain blocks only until this hash. This indicates that \texttt{end} is the highest known local block for this account currently.
        \end{enumerate}
        
    \item \textbf{bulk\_push} \\
    A $bulk\_push$ message is sent for each account that was determined to be out of sync at the remote peer to which we sent the $frontier\_req$. \\
    After a $bulk\_push$ message containing just the header is sent, a list of serialized blocks are sent over the same connection.
        
    \item \textbf{confirm\_req} \\
    Two $confirm_req$ messages are sent upon the detection of a fork, one for each of the conflicting blocks  \\
    This message consists of the header followed by the serialized block.
        
    \item \textbf{confirm\_acq} \\
    This message indicates a vote for a block. The weight of the vote is equal to the weight of the representative whose signature is present in this message.
    \textit{Fields}:
        \begin{enumerate}
            \item \texttt{sequence}: An integer indicating the current round of voting on this block.
            \item \texttt{block}: Block that is being voted on.
            \item \texttt{account}: Representative account that is generating this vote.
            \item \texttt{signature}: A confirmation that the vote is indeed generated by the said representative account.
        \end{enumerate}

\end{enumerate}

\section{Experiments}
\subsection{Keepalive messages}
\label{sec:nano-keepalive}
A peer is added to Nano's peer list if it sends a valid $keepalive$ message. Nano does not restrict the number of peers that can be added this way. Using UDP makes it hard to detect if the response $keepalive$ message was delivered to an open port. So we can flood Nano's peer table at a very low cost by spoofing IPs. These peers are removed only after the next \texttt{cutoff} period (5 minutes) because they haven't sent a $keepalive$ in this interval. This means we have to send a single UDP packet every 5 minutes for a  spoofed IP to persist in the peer list. On top of the node not being able to receive incoming messages in a timely fashion because of the flood, having a very large list of such peers means most of the outgoing messages that the node tries to send will fail because these IPs will make up most of the $\sqrt{peers}$ set. This makes this an effective DoS vector against a Nano node.

To illustrate how simple this attack is to execute, listing~\ref{lst:nano-attack} provides the complete Python function that can be run as is to flood fake $keepalive$ messages to a Nano node. It uses the \texttt{scapy} Python library to spoof UDP packets. 

\begin{lstlisting}[caption=Python function to flood keepalive packets, basicstyle=\small, language=Python, escapechar=\%, label={lst:nano-attack}]
def flood_keepalive(limit=42 * 10 ** 6, dst='localhost'):
    from scapy.all import RandIP, IP, UDP, send
    import struct
    port, magic_number = 7075, b'RC'
    v, type_, ext = 0x09, 2, 0
    payload = struct.pack('<2sBBBBH', magic_number, v, 
        v, v, type_, ext)
    for i in range(8):
        payload += struct.pack('<16s', b'42.42.42.42')
        payload += struct.pack('<H', 4242)

    src = RandIP()
    for i in range(limit):
        ip = IP(src=str(src), dst=dst)
        udp = UDP(sport=port, dport=port)
        spoofed_packet = ip / udp / payload
        send(spoofed_packet, iface="lo0", verbose=False)
\end{lstlisting}

Listing~\ref{lst:nano-attack-run} shows a single execution of this function against a locally running Nano node. This node was isolated and modified to drop outgoing packets. 1000 peers were added to the peer table in under 4 seconds.

\begin{lstlisting}[caption=Flood demo, basicstyle=\small, language=Python, escapechar=\#, label={lst:nano-attack-run}]
In [1]: from nano import RPCClient
In [2]: rpc = RPCClient('http://localhost:7076')

In [3]: rpc.version()
Out[3]: {'node_vendor': 'RaiBlocks 13.0','rpc_version': 1, 
'store_version': 10}

In [4]: len(rpc.peers())
Out[4]: 0

In [5]: %time flood_keepalive(limit=1000)
CPU times: user 2.68 s, sys: 505 ms, total: 3.19 s
Wall time: 3.72 s

In [6]: len(rpc.peers())
Out[6]: 1000
\end{lstlisting}

Since there is no limit on the number of peers in the table, memory usage of the node increases substantially under this attack. Table~\ref{tbl:flood-result} shows the increase in memory and CPU of the node under the flood of just one instance of the function. The node became unresponsive after which the test had to stopped.

\begin{table}[]
\centering
\caption{Keepalive flooding results}
\label{tbl:flood-result}
\begin{tabular}{|llll|}
\hline
\textbf{Time(seconds)}   & \textbf{Number of peers}     & \textbf{Memory (MB)}  & \textbf{CPU(\%)} \\
\hline
0   & 0     & 5.3 & 0.08\\
30  & 5262  & 13 & 2.8 \\
60  & 12008 & 22.8 & 6.4 \\
90  & 17353 & 30.4 & 9.5 \\
120 & 22309 & 41.7 & 12.6 \\
150 & 27790 & 50.8 & 16.6 \\
180 & 33696 & 58.5 & 20.5 \\
210 & 38046 & 70 & 24.7 \\
240 & 41500 & 77.1 & 27 \\
270 & 50955 & 89.5 & 34 \\
300 & 59467 & 105.8 & 39.6 \\
330 & 68190 & 120.5 & 48.2 \\
360 & 77217 & 136.1 & 53.6 \\
390 & 65450 & 133.9 & 82.8 \\
420 & 74034 & 132.8 & 90.1 \\
450 & 82802 & 147.8 & 126 \\
480 & 76119 & 142.3 & 136.4 \\
510 & 84928 & 150.5 & 142 \\
540 & 69847 & 157.9 & 175.6 \\
570 & 78281 & 153.9 & 186 \\
600 & 86785 & 149 & 191.6 \\
\hline
\end{tabular}
\end{table}

One possible fix for this issue would be a handshake with a randomly generated ID used in the ping/pong. This is already implemented in Ethereum.

\subsection{Frontier requests}
\label{sec:nano-frontier}
\begin{figure}
\caption{Frontier request statistics}
\begin{subfigure}[b]{0.5\textwidth}
\caption{Number of accounts fetched}
\label{img:frontier_total}
\includegraphics[width=\textwidth]{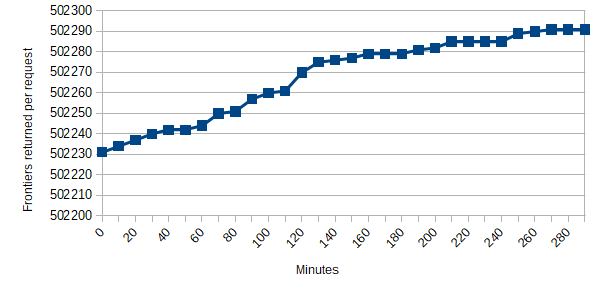}
\end{subfigure}
\begin{subfigure}[b]{0.5\textwidth}
\caption{Number of modified accounts}
\label{img:frontier_change}
\includegraphics[width=\textwidth]{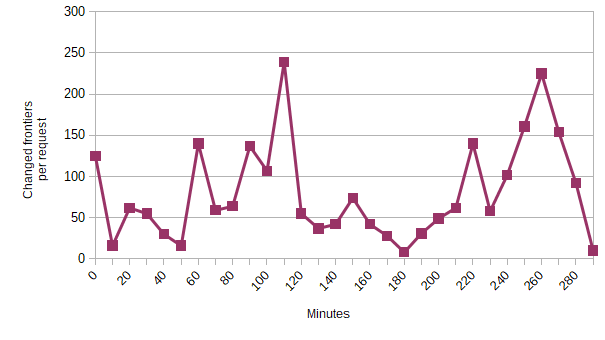}
\end{subfigure}
\end{figure}

A Nano node makes a frontier request to one of its peers once every 5 minutes. This is done to ensure that the latest block for each account is the same across both peers. The response to this request is the list of all accounts and their latest block hash. There are currently $\approx 550000$ accounts. Account and block hashes are 32-bytes each. So each node requests 32MB of data every 5 minutes. If a peer receives multiple frontier requests, its outbound bandwidth usage will be even higher. 
We collected responses to frontier requests from a production node for over 5 hours. Figure~\ref{img:frontier_total} shows the number of accounts returned for each request. But the actual number of accounts changing within each interval is very low as seen Figure~\ref{img:frontier_change}. This is because Nano has a low transaction rate at the moment. 

\begin{figure}
\caption{Frontier requests with IBLTs}
\begin{subfigure}[b]{0.5\textwidth}
\caption{Size of IBLT for each request}
\label{img:frontier_iblt_size}
\includegraphics[width=\textwidth]{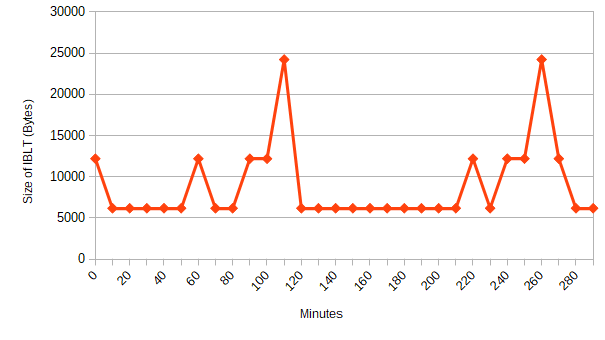}
\end{subfigure}
\begin{subfigure}[b]{0.5\textwidth}
\caption{Time taken to construct and transfer IBLTs}
\label{img:frontier_iblt_time}
\includegraphics[width=\textwidth]{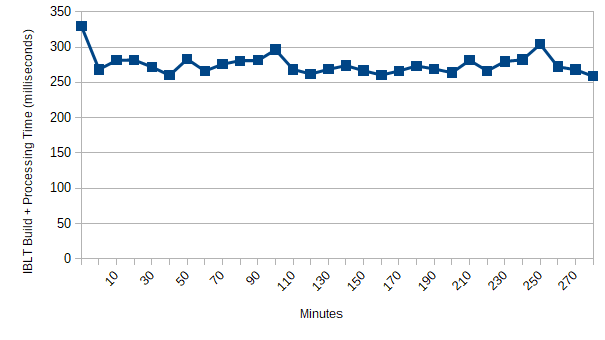}
\end{subfigure}
\end{figure}

This is a situation tailor-made for IBLTs because we have two sets whose difference is low in need of reconciliation. We inserted the collected data for each request into an IBLT and subtracted it from the IBLT constructed for the data from the previous request. The keys in the IBLT are the block hashes instead of the account hashes because most transactions change the frontiers of existing accounts. Accounts are stored in the value field of the IBLT. A new account is also reflected by a new frontier block and thus will be recovered along with the other block hashes in the new IBLT. We modified the size of the IBLT until all the new block hashes were recovered. Figure~\ref{img:frontier_iblt_size} shows the size of the IBLT constructed for each request. Note that the every frontier request at the moment downloads 32MB of data at the moment regardless of the number of accounts that may have changed since the previous request. The size of the IBLT reflects this delta. Only around $1MB$ of data would have transferred over 5 hours of frontier requests if IBLTs were in use. Instead each Nano node downloaded over $1.5GB$ of data in the same period. Each frontier request took $\approx$6 seconds to download all data. We measured the time it takes to build and transfer each IBLT (Figure~\ref{img:frontier_iblt_time}). This metric also improves to a few hundred milliseconds from 6 seconds. Multiparty set reconciliation described in ~\cite{multiparty} could be applied to this problem by expanding the scope of a frontier request to a subset of peers instead of a single peer. 

\chapter{Ethereum}
\section{Node Discovery}

Every node in ethereum has a private key. This key is separate from private keys associated with accounts that may be local to this node, for example the node's coinbase account if it is a miner\footnote{Private keys in Ethereum in are 32-bytes long and generated using the $secp256k1$ elliptic curve algorithm. The corresponding public key is 64-bytes because it is a point $(x, y)$ on the curve being used.} . The corresponding public key becomes this node's ID. This ID is then used to construct a node designator of the form \texttt{enode://<node ID>@<node IP>:<node Port>}. We have the option of using a different port for node discovery separate from the port used for general node-to-node communication. The discovery protocol uses UDP whereas node-to-node communication uses Ethereum's home-rolled RLPx on top of TCP. If these two ports are separate, the node designator will be of the form \texttt{enode://<node ID>@<node IP>:<TCP Port>?discport=<UDP Port>}.
 
The enode descriptors of bootstrap nodes are configured in code. Geth stores peer information from a session in a Level DB\cite{leveldb} key-value database. We have the option of specifying an on-disk file name for this database without which the database is maintained in memory. If a previous valid node database file is specified, at most 30 seed nodes that we communicated with in the last 5 days are picked at random and added to the bootstrap list. Bonding is then initiated with each node from the bootstrap list. It consists of sending a $Ping$ message with a randomly generated ID and waiting for a $Pong$ (see Networking Protocol for structure of these messages) with the same ID. The number of concurrent bondings in progress is limited to 16. On successful bonding, the peer is added to the routing table. A thread is now started to drop peers that are already present in our database from previous runs and haven't successfully bonded in the last 24 hours. This is to reduce the impact of dropping potentially useful seed nodes if the node is coming back to the network after a long time. This \textit{expirer} thread runs the drop once every hour.

The construction of the routing table and even the concept of node IDs is based on a peer-to-peer DHT protocol called Kademlia \cite{kad}. Kademlia uses a distance metric based on the $xor$ of node IDs to sort the routing table. Ethereum uses the 256-bit SHA3 hash of the node IDs instead of the 512-bit public key because the distribution of elliptic curve public keys is not uniform \cite{nonuniformec}. The table is divided into 256 equal-sized buckets. The size of each bucket is 16 nodes. A node is added to bucket $i$ if the position of the first 1 bit in the xor of  this node's ID and our ID is $i$. Since it is very unlikely for higher numbered buckets to be filled (the first $i - 1$ bits of two random SHA3 hashes have to match), Ethereum reduces the number of buckets to 17 with the last bucket acting as a catch-all for IDs with i from 17 to 255. Inside each bucket, nodes are sorted by when they were added. If a node is already present in a bucket, it is bumped to the front. If the bucket is full, a bonding is re-initiated with the least recently seen node in that bucket. The node which could not be added is placed in a replacement list to be chosen when a node is removed by failed bonding. There can be up to 10 replacement nodes for each bucket. A point to note here is that Ethereum does not actually use this table for the purpose it is meant for in Kademlia: to find nodes that store a particular key. No keys are being stored here. One possible rationale maybe to allow future scaling options such as sharding the blockchain.

The number of outbound connections to be established on start is determined by a $DialRatio$ which defaults to 3 and the maximum number of concurrent connections a node can have is set to 25. This means that we can connect to 8 peers at a time and the remaining slots are reserved for inbound connections. Before the routing table is looked up, the node tries to dial static and trusted nodes specified in the initial configuration. Outbound connections are established to trusted nodes even if the limit has been reached. Dialing involves performing an handshake involving the $Hello$ and $Status$ messages. If there are outbound slots remaining after attempting to handshake with static and trusted nodes, the routing table kicks into gear and we start by looking up a random node ID. Up to 32 nodes are fetched from buckets closest to this node ID and a $FindNode$ message containing this node ID is sent to each peer. By this time, we should already be present in the peer's routing table from a previous bonding because a peer is added to our routing table only after a successful bonding to start with.  The peer responds with a $Neighbors$ message containing 16 of its current peers that are closest to the sent ID. This message can be split into multiple UDP packets. After bonding with these peers they are inserted into the routing table with only 10 IPs from each /24 being allowed with a further restriction of at most 2 per bucket.

\section{Block Synchronization}

There are two ways for synchronization with a remote peer to start: if the number of peers reaches the desired minimum of 5 or by a force sync cycle that runs every 10 seconds. The peer with the highest known total difficulty is chosen as the best. This difficulty is retrieved during the handshake described above and also reset whenever a new block is received from a peer. Another condition for sync to begin is that the peer’s difficulty has to be higher than the difficulty of our latest block. If the node is being started for the first time, a genesis block is generated from a configuration embedded in code.\footnote{As an interesting aside, the $extra\_data$ field in the main network's genesis block was chosen to be the hash of a future block on the test network. Block number 1028201 was chosen because it is a palindrome and prime \cite{ethgenesis}. This was done so that every participant could generate the same block while still utilizing some randomness.} 

The best peer's head hash is fetched as a part of the $Status$ message. This header is now requested. Every time a request to a peer is made, a timer is set for the current TTL which is determined using the methods given in algorithm ~\ref{alg:fetch-ttl}. If there is an error or a timeout, this peer is dropped. If the header is fetched, the common ancestor block is determined between our local latest block and the remote head. The $floor$ limit for this search is $curLocalHead - 3 * proofOfWorkEpoch$. The current epoch is 30000 blocks and means that a new DAG is generated in the $ethash$\cite{ethash} mining algorithm that Ethereum uses. This means that a peer is not chosen for synchronization if it is on a fork that is separated by more than 90000 blocks from our own chain. To determine the ancestor,  we first start at ($curLocalHead - maxHeaderFetch$) and fetch headers at a step of 16. $maxHeaderFetch$ is set to 192 so we fetch a total of 12 headers. These headers are checked in the descending order of block number and the first header that is found on the local chain is considered as the ancestor. If this method fails, a binary search is initiated between the genesis block and $curLocalHead$.

\begin{algorithm}
    \caption{Fetch the current TTL threshold}
    \label{alg:fetch-ttl}
\begin{algorithmic}[1]
\State $rtt \gets 20$ \Comment Initialization
\State $rttConf \gets 1$ \Comment Confidence in our estimate
\State $ttlScaling \gets 3$
\State $ttlLimit \gets 60$
\State $qosTuningImpact \gets 0.25$
\Statex
\Procedure{TuneQOS}{} \Comment Runs as a separate thread every 20 seconds
\State $rtt \gets (1 - qosTuningImpact) * rtt + qosTuningImpact * \Call{GetPeerMedianRTT}{}$
\State $rttConf \gets rttConf + (1 - rttConf) / 2$ \Comment Increase confidence in our estimate
\EndProcedure
\Statex
\Procedure{GetTTL}{}
\State $ttl \gets ttlScaling * rtt / rttConf$
\If {ttl \textgreater ttlLimit}
\State \Return $ttlLimit$
\Else 
\State \Return $ttl$
\EndIf
\EndProcedure
\end{algorithmic}
\end{algorithm}

Concurrent processes are started to fetch headers, bodies and receipts in segments from different peers using this ancestor.  A receipt in Ethereum captures the logs generated by transactions which are calls to smart contract functions. As protection against peers sending bad headers, a skeleton is first built by requesting a subset of headers from the best peer.  The rule for this subset is to fetch every header at a step of $maxHeaderFetch$ on the peer’s chain starting from the ancestor until the number of headers in this subset is 128. Each of the header in this subset now becomes the start point for a header fetch from a different peer. Each fetch request is limited to $maxHeaderFetch$ headers.

After a segment of data is fetched from a peer, its throughput and RTT stats are updated. The median RTT obtained during the fetch is used to tune the TTL timer that we set for each request. This tuning is performed once every $RTT$ seconds.  The estimated $RTT$ is weighted by a confidence score which is decreased by a factor of $nPeers -1 / nPeers$ each time a new peer is added to our peer list . This decrease happens only if the number of peers is less than 10.

\section{Block Propagation}
When a new block is mined or received, it's hash is sent in a $NewBlockHashes$ message to all peers. The entire block is then sent to a square root of the number of peers.  

A new block message also includes the sender's total difficulty (TD) which is the cumulative sum of the difficulties of all the blocks preceding the current block. The receiver does not know if the included TD is truthful since calculating the true TD requires validating the entire chain up to this point and not every block in  the chain might be present yet. Hence, an estimate is made by summing of the new block's difficulty and the TD of its parent. This difficulty is used to pick the best peer for synchronization that we saw above. When the block is actually imported into the full chain, the calculated TD is compared to the advertised TD and the peer is dropped as malicious if there is a mismatch. Each peer also has a queue limit of 64 new blocks as an anti-DoS measure. The new block is also discarded if it is older than the current head by more than 7 blocks (uncle limit) or ahead by over 32 blocks.  Rebroadcast of the block happens as soon as the header is verified without the transactions themselves being checked. If the block competes against the current head block, it is stored away without processing until the total difficulty of this competing chain overtakes the canonical chain in which case this chain becomes canonical.

\section{Block Storage}

Ethereum uses the LevelDB key-value store for all of its data storage. This means that all of Ethereum's internal data structures have to be converted to a key/value format before being stored on disk. This is achieved using a custom serialization protocol called Recursive Length Prefix (RLP). A data object such as a block header, transaction, peer information is encoded into RLP. The $keccak256$ hash of this encoded data becomes the key and the RLP-encoded data itself is the value when inserted into LevelDB. Please refer to Appendix \ref{appendix:rlp} for examples of RLP encoding and decoding.

To efficiently store and retrieve the state of execution of smart contracts, Ethereum uses a trie data structure --- specifically a variant known as a Patricia Trie. This trie is combined with the properties of a general Merkle tree\cite{merkle} to provide cryptographic guarantees about the presence of data. Bitcoin uses Merkle trees to efficiently represent the transactions in a block. Only the root of the Merkle tree for a block is stored in the header. Any change to a transaction causes a change in the root thus changing the block header. Along with the root of the transaction trie, Ethereum also stores the root of the state trie and the receipts trie in the block header. Transaction tries and receipt tries are unique to a block whereas a state trie is a global data structure containing information about all contract accounts.

Smart contracts can be accessed on the blockchain using addresses of the same format as a normal currency-holding account. In the trie, the $kecccak256$ hash of this contract address forms the path from the root to the leaf which is the RLP encoding of the $Account$ object. This objects consists of fields for nonce (updated after every transaction involving this account to prevent replays), balance, hash of the smart contract code and a reference to the root of a separate trie which stores the values of variables associated with this smart contract. References to other tries or nodes within the same trie are just the keys for that corresponding node in the LevelDB database.  Each contract account has its own storage trie. 

A detailed walk-through of how these tries are constructed and a general explanation of Merkle Patricia trees is given in Appendix \ref{appendix:trie}.

Since LevelDB does not have an explicit concept of tables, Ethereum uses key prefixes to identify column classes. As an example, table~\ref{sch:eth-nodes} lists the mechanism used to store the routing table.

\begin{table}[]
\centering
\caption{Schema for routing table used in node discovery}
\label{sch:eth-nodes}
\begin{tabular}{|l|p{8cm}|}
\hline
\multicolumn{1}{|c|}{Key}                           & \multicolumn{1}{c|}{Value}                                                                         \\ \hline
\textless{}node\_id\textgreater + ":discover"       & RLP encoding of node details such as address, port, node ID and time of addition to routing table. \\ \hline
\textless{}node\_id\textgreater +  ":discover:ping" & Time of last ping sent                                                                             \\ \hline
\textless{}node\_id\textgreater + ":discover:pong"  & Time of last received pong received                                                                \\ \hline
\end{tabular}
\end{table}

Ethereum uses Bloom filters internally to record the presence of events in the logs generated for a transaction. As mentioned above, these logs are captured in the receipt for a block. Smart contracts can emit events while execution. These events can only be instances of registered $topic$s. Each $topic$ has a list of variables captured in the event. 

\section{Network Analysis}
Our crawler simulates a bonding and sends a $FindNodes$ message to extract peer information. The collected data was augmented and verified by multiple external sources\cite{ethnodes}. Over 19000 public nodes were detected.

\begin{figure}
\caption{Country-wise distribution of 19121 public Ethereum nodes}
\label{img:eth-country}
\includegraphics[scale=0.6]{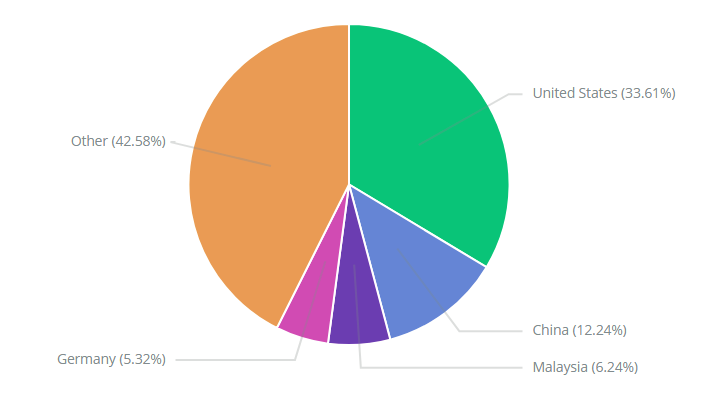}
\centering
\end{figure}

Figure~\ref{img:eth-country} shows a break down of public nodes by country. Ethereum's network is well decentralized with nodes spread out over 2000 Autonomous Systems as illustrated in Figure~\ref{img:eth-asn}. However, the top 5 ASNs still contain almost $30\%$ of the nodes and all of them belong to large cloud providers. Any outage at these services could mean a significant loss of hashing power as Ethereum still uses a Proof of Work based consensus scheme. (Table~\ref{tbl:eth-asn-table})

\begin{table}[]
\centering
\caption{Top 5 ASNs with public Ethereum nodes}
\label{tbl:eth-asn-table}
\begin{tabular}{|lp{7cm}l|}
\hline
\textbf{ASNum}   & \textbf{Name}     & \textbf{Percentage  of nodes} \\
\hline
16509 & AMAZON-02 - Amazon.com, Inc., US & 9.2\\
45102 & CNNIC-ALIBABA-CN-NET-AP Alibaba (China) Technology Co., Ltd., CN & 8.4 \\
14618 & AMAZON-AES - Amazon.com, Inc., US & 4.3 \\
14061 & DIGITALOCEAN-ASN - DigitalOcean, LLC, US & 3.4 \\
37963 & CNNIC-ALIBABA-CN-NET-AP Hangzhou Alibaba Advertising Co.,Ltd., CN & 3.3 \\
\hline
\end{tabular}
\end{table}

\begin{figure}
\caption{ASNs containing public Ethereum nodes}
\begin{subfigure}[b]{0.5\textwidth}
\caption{CDF}
\label{img:eth-asn}
\includegraphics[width=\textwidth]{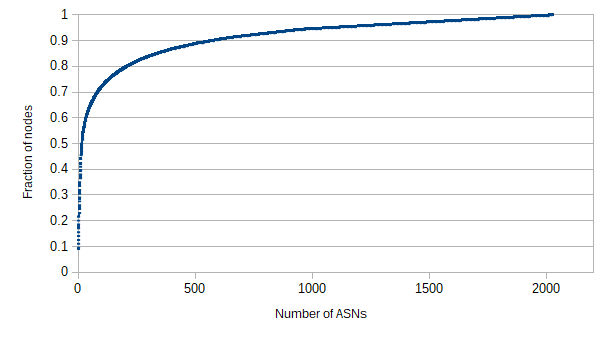}
\end{subfigure}
\begin{subfigure}[b]{0.5\textwidth}
\caption{Count}
\label{img:eth-asn}
\includegraphics[width=\textwidth]{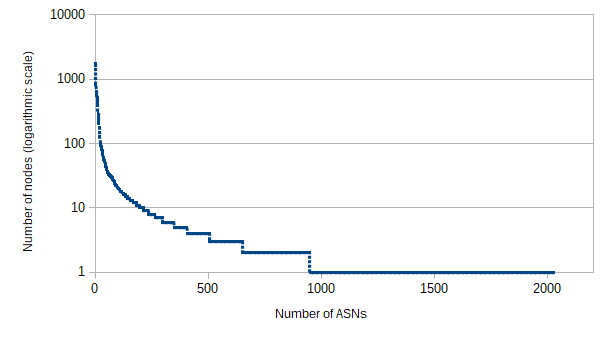}
\end{subfigure}
\end{figure}

\section{Networking Protocol}
Ping, Pong, FindNodes and Neighbors are messages are transmitted via UDP. The remaining messages use TCP.

\begin{enumerate}
    \item \textbf{Ping} \\
    \textit{Fields}:
        \begin{enumerate}
            \item \texttt{version}
            \item \texttt{from}
            \item \texttt{to}
            \item \texttt{expiration}
        \end{enumerate}
    
    \item \textbf{Pong} \\
    \textit{Fields}:
        \begin{enumerate}
            \item \texttt{to}: This field should be the same as the from address in the corresponding Ping packet. 
            \item \texttt{replytok}: Hash of the corresponding Ping packet.
            \item \texttt{expiration}: Timestamp after which this packet becomes invalid.
        \end{enumerate}
        
    \item \textbf{FindNodes} \\
    \textit{Fields}:
        \begin{enumerate}
            \item \texttt{target}: A 32-byte address which does not have to be an actual node ID (See Node Discovery for more details about Kademlia lookup)
            \item \texttt{expiration}
        \end{enumerate}
        
    \item \textbf{Neighbors} \\
    \textit{Fields}:
        \begin{enumerate}
            \item \texttt{nodes}: A list of real nodes IDs.
            \item \texttt{expiration}
        \end{enumerate}
        
    \item \textbf{Hello} \\
    \textit{Fields}:
        \begin{enumerate}
            \item \texttt{version}: Version of the P2P protocol to be used for communication
            \item \texttt{client\_version\_string}: Identify the node software (eg. geth, parity) similar to a User Agent string
            \item \texttt{capabilities}: Name and version of protocols supported. This is used to determine if any sub protocols such as the Light Ethereum Subclient are supported.
            \item \texttt{listen\_port}: Defaults to 30303
            \item \texttt{remote\_pubkey}: Our node ID
        \end{enumerate}

    \item \textbf{Status} \\
    \textit{Fields}:
        \begin{enumerate}
            \item \texttt{protocol\_version}: Subprotocol chosen from the \texttt{capabilities} sent in the previous $Hello$ message.
            \item \texttt{network\_id}: Identify main, test or private networks.
            \item \texttt{total\_difficulty}: Extracted from the latest block at the head of the chain
            \item \texttt{best\_hash}: Hash of the latest block
            \item \texttt{genesis\_hash}
        \end{enumerate}
 
      \item \textbf{GetBlockHeaders} \\
       \textit{Fields}:
        \begin{enumerate}
            \item \texttt{block\_number\_or\_hash}: Get all headers from this hash or block number
            \item \texttt{max\_headers}: Do not fetch more than this number of headers 
            \item \texttt{skip}: Block number to skip
            \item \texttt{reverse}:Send latest headers first
        \end{enumerate}
        
         \item \textbf{BlockHeaders} \\
       \textit{Fields}:
        \begin{enumerate}
            \item An RLP-encoded list of block headers where each header has the following fields:
            \begin {enumerate}
                \item \texttt{parent\_hash}
                \item \texttt{uncles\_hash}: Combined hash of the headers of all uncle blocks of this block
                \item \texttt{coinbase}: Address of the miner
                \item \texttt{state\_root}: Hash of the root of the state trie after this block has been imported (See Block Storage for details)
                \item \texttt{receipt\_root}
                \item \texttt{transaction\_root}
                \item \texttt{bloom}: Used to verify the presence of logs in the receipt trie.
                \item \texttt{difficulty}
                \item \texttt{block\_number}
                \item \texttt{gas\_limit}
                \item \texttt{gas\_used}
                \item \texttt{timestamp}
                \item \texttt{extra\_data}
                \item \texttt{mix\_hash}: Generated from the DAG used by ethash. Used in difficulty verification.
                \item \texttt{nonce}
            \end{enumerate}
        \end{enumerate}
        
         \item \textbf{GetBlockBodies} \\
       \textit{Fields}:
        \begin{enumerate}
            \item \texttt{hashes}: An RLP encoded stream of block hashes whose bodies are being requested.
        \end{enumerate}
        
         \item \textbf{BlockBodies} \\
       \textit{Fields}: \\
        A list of RLP encoded block bodies. Each block body is itself composed of a list of transactions and a list of uncle headers. 
\end{enumerate}

\section{Experiments}
\subsection{Blockchain Analysis}
\begin{figure}
\caption{Transaction count distribution across $\approx$5.5 million blocks}
\label{img:eth-histogram-all}
\includegraphics[scale=0.7]{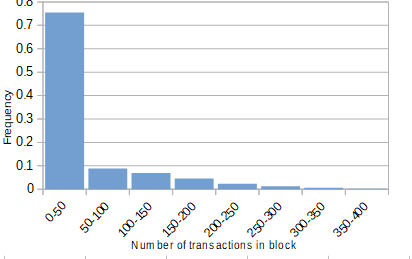}
\centering
\end{figure}

\begin{figure}
\caption{Transaction count distribution across $\approx$700000 blocks in 2018}
\label{img:eth-histogram-2018}
\includegraphics[scale=0.7]{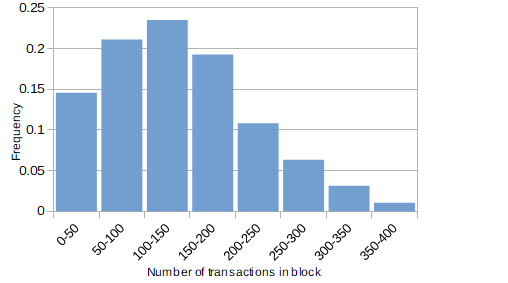}
\centering
\end{figure}

To estimate the amount of duplicate transaction data being transferred during block propagation, we analysed
more than 5.5 million blocks starting from Ethereum's genesis block in 2015 to blocks up to April 29 2018 were analysed. Figure~\ref{img:eth-histogram-all} shows the distribution of number of transactions in a block. Over 75\% of blocks historically have had no transactions in them. Things improve when only blocks from 2018 are considered (Figure ~\ref{img:eth-histogram-2018}) where there is an almost even distribution between 0-150 transactions.

\begin{figure}
\caption{Average transaction size and count over time}
\label{img:eth-size-count-time}
\includegraphics[scale=0.5]{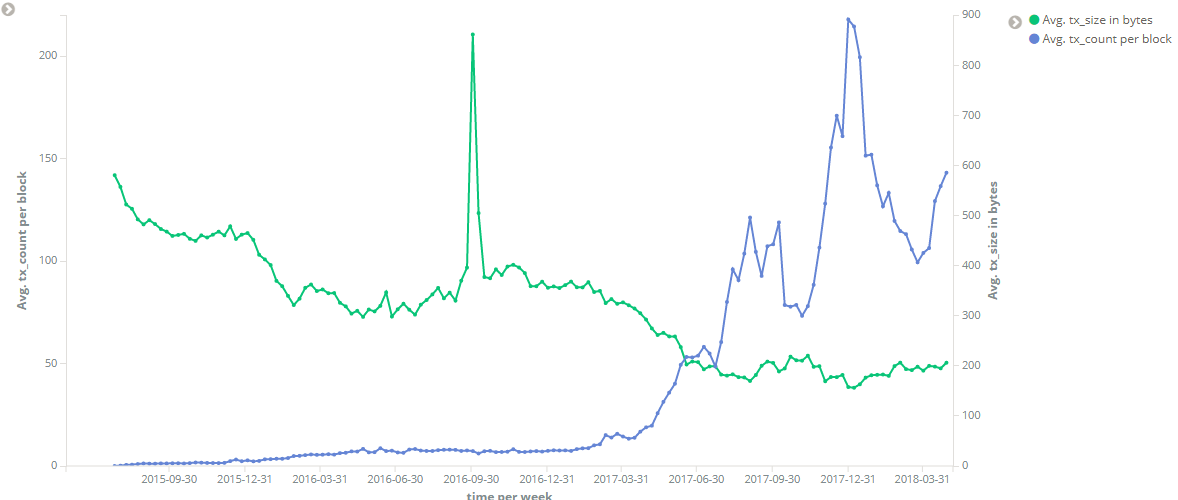}
\centering
\end{figure}

\begin{figure}
\caption{Share of Smart Contract calls}
\label{img:eth_smart_contract}
\includegraphics[scale=0.7]{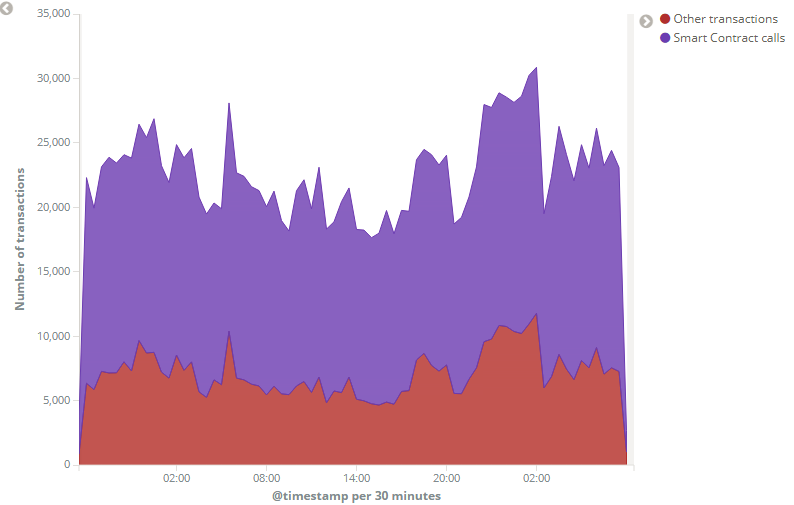}
\centering
\end{figure}

Figure~\ref{img:eth-size-count-time} shows that though the number of transactions per block has stabilized, the average size per transaction continues to grow. This is because a larger number of transactions are calls to Smart Contract functions which include input data. Figure ~\ref{img:eth_smart_contract} shows the distribution between Smart Contract function calls and regular value transfer transactions in blocks between April $29^{th}$ and April $30^{th}$. There are on average 5 Smart Contract calls for every regular transaction. The \texttt{input} field in Smart Contract calls has an average size of $\approx 311$ bytes. This field is empty for regular transactions.

\subsection{Graphene in Geth}
\label{sec:eth-experiment}

We first describe the normal course of propagation of a block in Geth before looking at the changes made for Graphene. When a new block is mined, it is first propagated in full to $\sqrt{peers}$ via a \texttt{NewBlockMsg}. A full block includes the block header, list of transactions and list of uncle block headers. The block hash is then announced to all peers via a \texttt{NewBlockHashesMsg}. Upon receiving a \texttt{NewBlockHashesMsg} a peer sends out separate request for headers and body via \texttt{GetBlockHeaders} and \texttt{GetBlockBodies} messages. Once a full block has been fetched and only its header verified, it is pushed in full to $\sqrt{peers}$. After the block transactions are verified and the block is ready to be imported into the blockchain, its hash is announced to all peers. 

The implementation for Graphene starts by disabling full block broadcasts. Only \texttt{NewBlockHashesMsg}s are sent. The response to these messages are 
  \texttt{GetGrapheneMsg} messages which contain the number of pending transactions at the receiver. The sender now constructs a \texttt{GrapheneMsg} containing a Bloom filter and an IBLT. The sender also sends  This message also includes the number of transactions in the block which is used to verify transaction recovery and the number of cells in the IBLT which is needed for the receiver to construct their own IBLT. Only the first five bytes of the transaction hash are inserted into the Bloom filter and IBLT. This provides a reasonable degree of randomness while reducing the number of bytes required.  To recover transaction indices, we have implemented the lexicographic sort technique mentioned in the Graphene paper. To do this, the sender sorts the extracted short transaction IDs. This list is used to construct an array of integers indicating the index of the transaction. At the receiver's end, after the extra transactions added because of the Bloom Filter's $fpr$ are detected by the IBLT, we check if the remaining transactions that passed through the Bloom Filter equal the number of transactions in the block. If the number is greater, it means that some transactions from the block are missing from the receivers pending list and we resend a \texttt{GetGrapheMsg} by doubling the size of the included pending list. If remote transaction IDs are recovered from the IBLT subtraction, we send a \texttt{GetTxMsg} for these IDs. This message does not exist in Ethereum's RPC protocol and was implemented newly for Graphene. Lastly, the \texttt{GrapheneMsg} message also includes the list of uncle block headers. This provides the receiver sufficient information to reassemble the list of transactions and uncle block headers and queue the block for validation and import into the blockchain.
 
\begin{figure}
\caption{Block propagation sizes for 50 blocks starting at block 5531100}
\label{img:geth_graphene_size}
\includegraphics[scale=0.7]{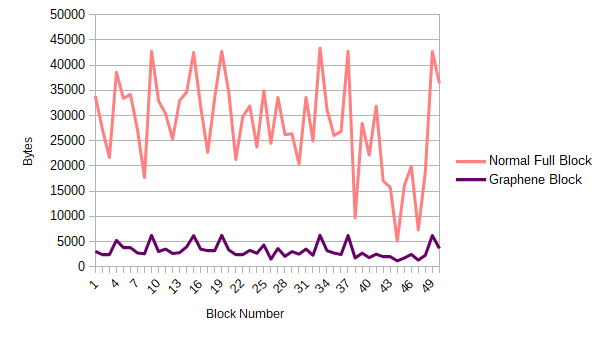}
\centering
\end{figure}

 We ran the implementation against 50 blocks from April 29, 2018. The size of the mempool was kept constant at 60000. An analysis of historical pending transaction count~\cite{eth-pending} shows that this is a comfortable upper bound. To simulate the presence of 60000 transactions in the pending list, an array of random 5-byte strings was prepopulated. After real transactions from the pending list were tested against the Bloom Filter and inserted into the IBLT, this array was iterated over.  Figure~\ref{img:geth_graphene_size} shows the sizes of RLP encoding of the blocks which was actually transmitted across the wire. Thus, we account for all serialization overheads added by the different data structures. We observe an 8x improvement on average.

\begin{table}[]
\centering
\caption{Average processing time for block propagation messages}
\label{tbl:geth_processing_time}
\begin{tabular}{|llll|}
\hline
\textbf{GetBlockBodiesMsg}   & \textbf{GetGrapheneMsg}     & \textbf{BlockBodiesMsg}  & \textbf{GrapheneMsg} \\
\hline
26.09$\mu$s & 1.39ms & 1.171ms & 14.04ms \\
\end{tabular}
\end{table}

\begin{table}[]
\centering
\caption{Breakdown of \texttt{GrapheneMsg} processing}
\label{tbl:geth_breakdown}
\begin{tabular}{|ll|}
\hline
\textbf{Deserialize received data}   & 135.08$\mu$s \\
\textbf{Test pending list against Bloom filter} & 13.91ms \\
\textbf{Decode IBLT}  & 41.32$\mu$s\\ 
\textbf{Sort and extract transaction IDs} & 128.63$\mu$s \\
\hline
\end{tabular}
\end{table}

We also tracked the impact of building and checking the Bloom Filters and IBLTs. Table~\ref{tbl:geth_processing_time} shows the average time taken comparing corresponding messages from the normal Ethereum protocol against ones implemented by Graphene. Though the increase might seem substantial, the times posted should still be very fast in the context of normal operations of an Ethereum node.  Because of the cost involved in checked the size of the pending list, the average time taken for \texttt{NewBlockHashesMsg} also went up from 29.24$\mu$s to 131.92$\mu$s. Table~\ref{tbl:geth_breakdown} provides a detailed listing of the times taken by different components involved in processing a \texttt{GrapheneMsg}. Almost all of the huge time increase is because of the cost of checking over 60K keys in the Bloom Filter.

\chapter{IOTA}
\section{Node Discovery}
 
IOTA does not have an automatic node discovery mechanism. Neighbors have to be manually configured either at start or added via the $addNeighbors$ HTTP API. To successfully communicate with a neighbor, the peer must also run the $addNeighbors$ command locally on their end to add our address and port to their valid neighbors list. No other form of handshake or validation occurs before messages are transmitted to and from this neighbor. The project developers recommend finding peers in dedicated channels on sites like Discord or Reddit. However, there is a third-party project called Nelson which provides automated peer management for IOTA. It does this by accessing the $addNeighbors$ and $removeNeighbors$ HTTP APIs locally and communicating with other peers running Nelson via its own HTTP messages.
 
If no neighbors are listed when starting Nelson, it fetches a static list of entry nodes from Github\cite{nelson-boot}. This initial list of neighbors is considered trusted and priority is given to these nodes when connecting or advertising to other Nelson instances by setting their weight to 1.  A websocket connection is established to each of the neighbors with details about the IOTA node (fetched via the $get\_node\_info$ API) passed in the header. Upon receiving a websocket connection, the neighboring Nelson instance will check if the two IOTA nodes are compatible (both nodes should agree on using either UDP or TCP for communication). If the connection is valid, the requesting node is added as a peer with an initial weight of 0. The neighbor then responds on the same websocket connection with a list of up to 6 of its neighbors sorted by weight, peer quality and connection age (see algorithm~\ref{alg:nelson-peer-trust}). The health of the IOTA node is checked every 15 seconds via the $get\_node\_info$ and $get\_neighbors$ IOTA APIs. The $get\_neighbors$ API reports the number of random transaction requests, new transactions and invalid transactions per active peer. This information is used by Nelson to determine the peer quality.

\begin{algorithm}
    \caption{Peer trust calculation in Nelson}
    \label{alg:nelson-peer-trust}
\begin{algorithmic}[1]

\Procedure{GetPeerQuality}{$invalidTxs, randomTxRequests, newTxs$}
\State $badTxThreshold \gets 3$
\State $badTxs \gets invalidTxs + randomTxRequests$
\State $badTxRatio \gets (invalidTxs * 5 + randomTxRequests) / newTxs + 1$
\If {not \Call{istrusted}{$peer$} and $newTxs = 0$ and $badTxs > badTxThreshold$}
\State $penalty \gets 1 / (badTxs + 1)$
\Else
\State $penalty \gets 1$
\EndIf
\State \Return $(1 / badTxRatio + 1) * penalty$
\EndProcedure
\Statex
\State $peerScore \gets connectionAge * \Call{getpeerquality}{get\_node\_info(peer)} * (1 + peerWeight * 10)$
        	
\end{algorithmic}
\end{algorithm}

Peer connections are managed in terms of epochs with the default epoch interval being 1200 seconds. Every time a new epoch is started, a random number of peers are dropped and connections are re-established anew with our node assuming a new identity. Within each epoch, peers are periodically checked for health in every cycle with the default cycle interval set to 60 seconds. If a peer has not sent new transactions in the last 5 minutes, it is considered lazy and dropped. The IOTA node itself reruns DNS resolution for all its neighbors every 180 seconds.
 
\section{Block Synchronization}
\label{sec:iota-sync}
 
Since IOTA has no explicit concept of a block, let's take a look at how transactions are linked together which will then let us understand how transactions are requested and transmitted.
 
In a nutshell, IOTA works by having each transaction approve two previous transactions by including their hashes in its signature. Unapproved transactions are called tips. This tree-like data structure is called tangle.  A transaction’s confirmation confidence is given by the number of times a valid tip selected by the tip selection algorithm approves it either directly or indirectly. The tip selection algorithm is a weighted random walk starting from the genesis transaction. The weights ensure that `lazy' tips --- tips that approve old transactions and thus keep the tangle from moving forward --- are discouraged. The cumulative weight of a transaction is the count of the number of approvers it has.
 
Transactions are considered confirmed with a confidence of 100\% if they are approved by a milestone transaction which is generated every minute by a special account known as the coordinator. The coordinator’s address is hardcoded and controlled by the IOTA developers. The current plan is to keep the coordinator functioning until the network has enough vested participants to generate steady, valid traffic. Like Nano, all of the IOTA that can ever exist was created in the genesis transaction $(3^{33}-1/2)$.
 
To keep the transaction size small enough to fit into a single datagram packet, a transaction is divided into sub-transactions all of which belong to the same bundle.  There are four types of sub-transactions and all sub-transactions of the same type have the same index.  Index 0 contains sub-transactions which specify the receiving accounts and values for this transaction. Index 1 sub-transactions specify the sending addresses and the first part of transaction signatures. The value of these transactions is negative because coins are being spent. The second half of the signatures are present in index 2 sub-transactions. Index 3 sub-transactions specify the change addresses to which remaining balance should be sent if input is greater than output. All sub-transactions share the same bundle hash separate from the transaction hashes that uniquely identify each transaction.
 
All transactions in a bundle except the last contain the same branch hash. Branch is the name given to one of the two previous transactions that each transaction has to approve. This hash belongs to a random transaction not in this bundle and has to be a tip. A trunk transaction is the second transaction approved by a transaction. Trunk transactions are used to chain transactions inside a bundle together. A sub-transaction at index 0 will have the hash of a sub-transaction at index 1 in its trunk field. For the last transaction in the bundle, the branch transaction of the previous transactions become the trunk and the branch transaction is a new random tip. Each bundle in effect approves two external transactions.
 
If a transaction’s confirmation confidence does not increase in a reasonable amount of time because of almost similar cumulative weights during tip selection and the algorithm’s preference for newer tips, the entire bundle can be reattached to the tangle. This requires replaying the transactions by regenerating the proof of work included in each transaction. This changes the transaction hash but the bundle hash remains the same and the node can detect the reattachment and discard the old attachments. If the reattached transactions are still consistent (valid path to genesis), the promote API generates zero-value transactions which will approve this transaction and a milestone transaction as a way of increasing the confirmation confidence.
 
A snapshot text file containing account balances is generated periodically by the IOTA developers and bundled with the code. On boot, this file’s signature is validated with a hard coded public key and the internal state of accounts is updated. The corresponding milestone transaction number provided is used as the latest known milestone index.
 
IOTA does not differentiate the first time a node starts from the general operation of a node. There is no special download of transactions. After the snapshot is loaded, the procedure described in the Block Propagation section kicks in.
 
\section{Block Propagation}
 
A $tipsRequestingPacket$ message is sent to all neighbors every 5 seconds. This request contains the latest known milestone transaction's hash appended as a request to the hashes of the bundle, branch and trunk transactions and the receiver address of the same milestone transaction. This is because all messages in IOTA piggyback a request for another transaction when themselves sending a transaction. In this case, since the requested transaction hash is the same as the sent transaction, the receiver responds with a random tip transaction with a probability of 0.66. This happens only if the receiver itself has transactions to be requested in which case it responds with a $sendingPacket$ message containing the random tip and a random transaction from its request queue. The transaction being requested will be a milestone transaction with a probability of 0.7. Thus, a $sendingPacket$ has a different requested transaction and included transaction whereas a $tipsRequestingPacket$ has the same for both.
 
If a node does not have the transaction requested in the $sendingPacket$ or $tipsRequestingPacket$ message it is added to its request queue with a probability of 0.01. If transaction in the message is valid, the receiver broadcasts it to all its neighbors. The transaction is considered solid if both its branch and trunk transactions are themselves solid. If either the branch or trunk is not present, it is added to the request queue and the transaction is not considered solid. When an entire sub tangle becomes solid, the snapshot is updated to reflect the new balances for the accounts in the transactions in this sub tangle.
  
\section{Block Storage}
 
IOTA uses RocksDB\cite{rocksdb}, a Facebook derivative of Google's LevelDB, the key-value store we've seen being used in Ethereum and also used in Bitcoin. RocksDB offers support for column families for the key-value pairs. We've seen application-level schemas being defined in other projects which then transform to prefixes for fixed-format string keys while writing to the key-value stores. RocksDB provides this logical partitioning at the persistence layer itself. Two separate column families can be written to concurrently while still maintaining atomicity. Each column family can be configured separately, for example with different cache and compaction settings.  In addition RocksDB provides improved performance by using Bloom filters, support for transactions among other improvements and faster in-memory data store\cite{rocksdbvs}.

IOTA uses the following column families:
\texttt{tag, bundle, approvee, address, stateDiff,milestone, transaction-metadata, transaction}

 In addition, IOTA also has the functionality to publish to a message queue running ZeroMQ\cite{zmq} or exporting the blockchain to a file while the node is in operation.
 
\section{Network Analysis}
Using Nelson's information as the base, information about 470 IOTA's nodes was collected. This data was verified and augmented by data collected from \cite{iotanodes}.

\begin{figure}
\caption{ASNs containing public IOTA nodes}
\begin{subfigure}[b]{0.5\textwidth}
\caption{CDF}
\label{img:iota_asn_cdf}
\includegraphics[width=\textwidth]{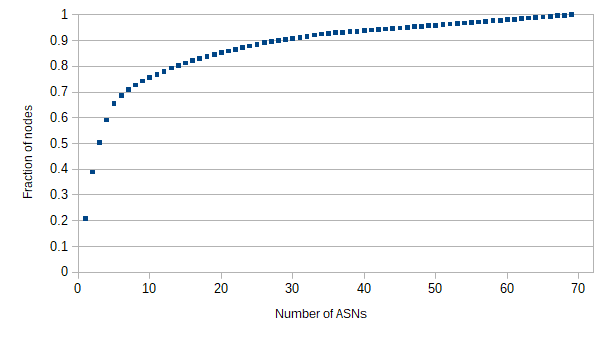}
\end{subfigure}
\begin{subfigure}[b]{0.5\textwidth}
\caption{Count}
\label{img:iota_asn_hist}
\includegraphics[width=\textwidth]{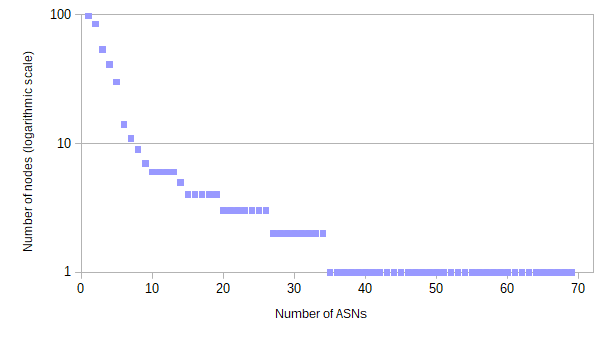}
\end{subfigure}
\end{figure}

IOTA's network is extremely centralized with the top 3 ASNs containing more than half of the nodes (Figures~\ref{img:iota_asn_cdf} and ~\ref{img:iota_asn_hist}).

A curiosity to note here is that ASN 14061 has the highest number of Nano nodes and the second highest number of IOTA nodes.
\begin{table}[]
\centering
\caption{Top 5 ASNs with public IOTA nodes}
\label{tbl:iota-asn-table}
\begin{tabular}{|lp{7cm}l|}
\hline
\textbf{ASNum}   & \textbf{Name}     & \textbf{Percentage  of nodes} \\
\hline
51167 & CONTABO - Contabo GmbH & 20.9 \\
14061 & DIGITALOCEAN-ASN - DigitalOcean & 18.1 \\
24940 & HETZNER-AS - Hetzner Online GmbH & 11.5 \\
197540 & NETCUP-AS - netcup GmbH & 8.7 \\
15169 & AS - GOOGLE & 6.4 \\
\hline
\end{tabular}
\end{table}

\begin{figure}
\caption{Country-wise distribution of 470 public IOTA nodes}
\label{img:iota-country}
\includegraphics[scale=0.6]{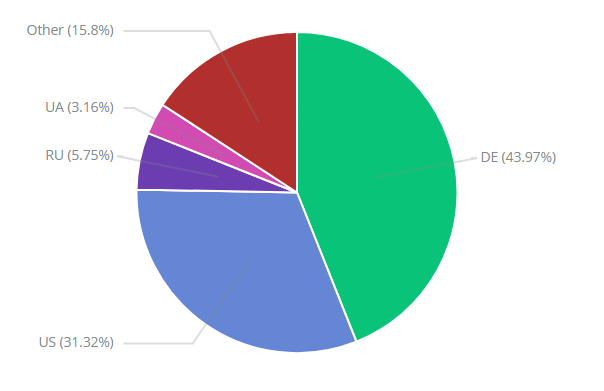}
\centering
\end{figure}

\section{Networking Protocol}
 
 The two message types in IOTA $tipsRequestingPacket$ and $sendingPacket$ are datagram packets of size 1650 bytes by default. They consist of a serialized transaction and a request for another transaction.

\section{Experiments}
\label{sec:iota-experiment}
IOTA uses the solidity of transactions to determine if a sub-tangle headed by a milestone can be used to update the account balance snapshot. This means that when a tip arrives, its branch and trunk transactions are checked for solidity. This involves fetching the transaction from disk and testing its $isSolid$ attribute.  Instead, we could use an in-memory Cuckoo filter to speed up checks for non-solid transactions. A Cuckoo filter is a probabilistic data structure akin to Bloom filters but provides constant time lookups and inserts and more importantly, constant time delete and list operations. It also has better space efficiency when the desired the false positive rate is less than 0.03. IOTA via RocksDB also has the option of using a Bloom filter for a sorted static table (SST) storing a column family but it is currently unused. Even if it is used, supporting deletions would require the use of Bloom filter variants like Counting Bloom filters\cite{countingbloom} which have much worse operation efficiency than Cuckoo filters. 

Our algorithm works as follows: All non-solid transactions are inserted into a Cuckoo filter. When a tip arrives, it's branch and trunk transactions can be first checked in the filter. If they are not present, we can be definitely sure that these transactions are already solid. This saves a disk access and since the number of non-solid transactions is bounded under normal operation, the size of the filter can be small. Once a transaction becomes solid, it is deleted from the filter. 

\begin{figure}
\caption{Simulation of insertion in probabilistic filters}
\label{img:cuckoo-insert}
\includegraphics[scale=0.8]{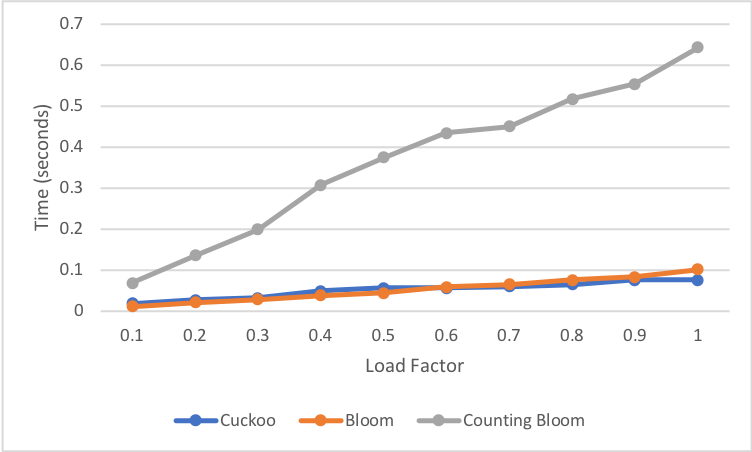}
\centering
\end{figure}

\begin{figure}
\caption{Simulation of access in probabilistic filters}
\label{img:cuckoo-access}
\includegraphics[scale=0.8]{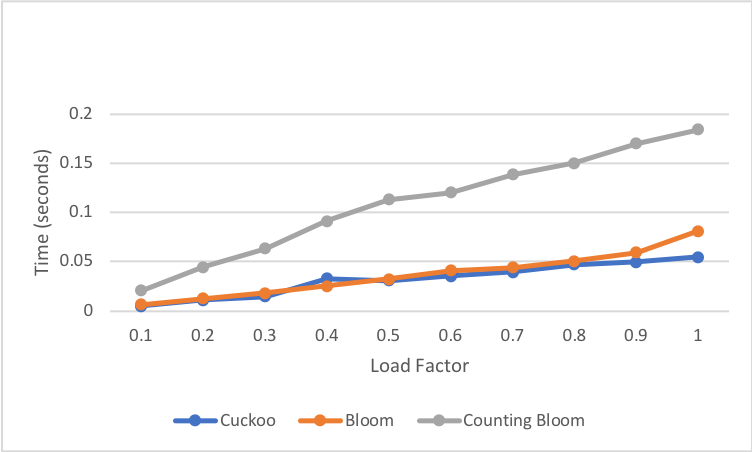}
\centering
\end{figure}

Figures~\ref{img:cuckoo-insert} and ~\ref{img:cuckoo-access} show the results of a simplistic Python simulation of the different filters. The results for Cuckoo filter could be much better with optimization but they already have the huge advantage of supporting deletion over Bloom filters with comparable times.

%\chapter{Bitcoin}
%\import{./}{bitcoin.tex}

% The following produces a numbered bibliography where the numbers
% correspond to the \cite commands in the text.
\specialhead{LITERATURE CITED}
\begin{singlespace}

\end{singlespace}

%%%%%%%%%%%%%%%%%%%%%%%  For Appendices  %%%%%%%%%%%%%%%%%%%
\appendix    % This command is used only once!
\addtocontents{toc}{\parindent0pt\vskip12pt APPENDICES} %toc entry, no page #
\chapter{Graphene} \label{appendix:graphene}
We first present a general description of Graphene's core components followed by a listing of Graphene's protocol for block transfer.

\section{Bloom filters}
A Bloom filter \cite{bloom} is a probabilistic data structure used to efficiently answer set membership queries. It has no false negatives and a configurable false positive rate.

Let $S = s_1, s_2, s_3, ..., s_n$ be a subset of strings from the universe $U$ of all possible strings such that $U >> m$.
Bloom filter $B=b_1, b_2, ..., b_m$ is a bit-vector of length $m$ which can determine if a string $s$ is in set $S$. A `yes' answer \textit{may} mean $s$ is in $S$. A no answer definitely means that $s$ is not in $S$. We do this by choosing $k$ independent hash functions $h_1, h_2, ..., h_k$. The false positive rate for a Bloom filter can be obtained as follows:

The probability that a given bit $l$ is not set after the inserting all $m$ element is:

$$ (1 - \frac{1}{m})^{kn} = e^{-\frac{kn}{m}}$$

The false positive rate is then simply the probability that a given bit $l$ is set after $m$ insertions by $k$ hash functions:
$$f = (1 - (1 - \frac{1}{m})^{kn})^{k} = (1 - e^{-\frac{kn}{m}})^{k}$$

By trying to minimize the derivate of the equation, we obtain the optimum value for $k$, the number of hash functions:
$$ k = ln 2 * \frac{m}{n}$$

\section{Invertible Bloom Lookup Table}

An Invertible Bloom Lookup Table is a data structure that can store key-value pairs. In a fashion similar to Bloom filters, each key-value pair is stored in k cells determined by k independent hash functions.  Each cell has a $keySum$, $valueSum$ and $count$ field.  When multiple key-value pairs map to the same cell, their keys and values are XOR’ed to form a single sum (hence the $-sum$ in the field names). The count field tracks the number of key-value pairs in a cell. Insertions, deletions and lookups can be performed on an IBLT in $O(k)$ time. Insertions and deletions always succeed. Key lookup and listing of all the key-value pairs are probabilistic and succeed with a high probability if the number of pairs n in the IBLT is less than a threshold parameter $t$.  For an explanation about how this parameter is chosen, please see \cite{iblt-paper}.

\section{Set Reconciliation using IBLT}

We now see how an IBLT can be used for set reconciliation. As described before, set reconciliation is the process of symmetrically synchronizing two identical sets of elements which differ only by a few elements. IBLT achieves this with a low overhead.
Alice and Bob are in possession of sets $S_A$ and $S_B$ respectively which they wish to reconcile.
They each build an IBLT T separately by inserting elements from their sets using hash functions $h_1,\ldots, h_k$. Bob sends his IBLT over to Alice. Note that the number of the cells in the IBLT is only dependent on the size of the symmetric difference d between the two sets and not the number of elements. Therefore there needs to be a prior round of communication to infer or calculate $d$.

Alice subtracts Bob’s IBLT from her own IBLT. Subtracting an IBLT consists of the following steps:
\begin{enumerate}
\item Find a \textit{pure} cell with count 1. The data in this cell is an actual set element $(K, V)$ not combined with any other elements via an XOR sum.
\item Delete this element from all cells $T[h_{1}(K)], T[h_{2}(K)]...T[h_{k}(K)]$ in which $(K, V)$ was originally inserted. The operation to delete a key is the same as insertion i.e we XOR the key to the current value in the cell which has the effect of removing that key from the cell.
\item If the above delete operation produced more \textit{pure} cells, repeat Step 2 for the element in this cell.
\item A key detail here is that a count of 1 does not guarantee a clean deletion of all other keys. For example, consider cell C which contains the XOR of 4 elements:
	$$ C = a \oplus b \oplus c \oplus d $$
C now has the count 4. To extract the value d, we need to delete a, b and c successively from C after which its count is 1. But if instead of c, some other element e was deleted from this cell, the count reduces to 1 while the value left behind is
$$	C = d \oplus e $$
To tackle this, IBLTs include another field known as the $hashKeySum$ field. Before insertion, each key is hashed using a separate hash function H and added to this field. The content of a \textit{pure} field $(K, V)$ is valid only if $H(K)$ equals the corresponding value in the $hashKeySum$ field. 
\end{enumerate}

All cells left behind with counts of 1 or -1 contain keys that were added or missing from Alice’s set respectively.

\section{Block Propagation using Graphene}

Graphene relies on the Set Reconciliation method described above to determine which transactions belong to a block. This is done without transferring the explicit transaction hashes. The steps involved in propagating a Bitcoin block using the Graphene protocol are:

\begin{enumerate}
\item A node in possession of a new block send an $inv$ to all of its peers. 
\item A peer responds with a $getgraphene$ message. This message includes the size m of the peer’s mempool.  This value is necessary to calculate the size of the symmetric difference as mentioned above.
\item The sender node sends a $graphene$ message containing a Bloom filter $S$ and IBLT $I$.  
\item The receiving peer creates an IBLT $I^\prime$ from the transactions in its mempool which are present in $S$.  Because $S$ can have false positives, $I^\prime$ can contain transactions not in the block. 
\item To further filter out these false positives, $I$ is subtracted from $I^\prime$ to extract the extraneous transaction IDs. This also gives us the IDs of any transaction included in the block but not present in the receiver’s mempool. Separate $getdata$ messages are sent for each such transaction.
\end{enumerate}

\chapter{RLP} \label{appendix:rlp}
Recursive Length Prefix (RLP) is Ethereum's homegrown serialization protocol. It boils data down to only two distinct types: a string (sequence of bytes) and a list of items (strings or nested lists). 

By just looking at the first byte of an RLP-encoded stream, we can gain sufficient information about the type and length of the data that follows. Let's look at the ranges of values for this byte and what they represent using the $rlp$ python library\cite{pyrlp}:

\begin{enumerate}
\item \textbf{0 to 127 ($0x00$ to $0x7f$)}: This value represents the actual data without any serialization overhead being added. This data is a number less than 128 and its interpretation is left to the process deserializing this stream i.e it may be treated as an ASCII character or an integer.

\begin{lstlisting}[caption=RLP encoding of a single byte, language=python, escapechar=\%, label={lst:rlp1}, showstringspaces=false]
In [1]: rlp.encode(1)
Out[1]: b'\x01'

In [2]: rlp.encode(127)
Out[2]: b'\x7f'

In [3]: # 0 falls under the next case
In [3]: rlp.encode(0)
Out[3]: b'\x80'
\end{lstlisting}

\item \textbf{128 to 183 ($0x80$ to $0xb7$)}: The RLP byte now indicates that the data that follows is a string of length between 0 and 55 bytes. The length of the string can be obtained by subtracting 128 from the value of this byte (which is the encoding for 0 as seen above). This byte is then followed by the actual string.

\begin{lstlisting}[caption=RLP encoding of 0-55 byte strings, language=python, escapechar=\%, label={lst:rlp2}]
In [5]: rlp.encode('')
Out[5]: b'\x80'

In [6]: rlp.encode('a' * 50)
Out[6]: b'\xb2aaa...aaa'

In [7]: 0xb2
Out[7]: 178

In [8]: 178 - 128
Out[8]: 50
\end{lstlisting}

\item \textbf{184 to 191 ($0xb8$ to $0xbf$)}: Subtract this number by 183, let's call this value $s$. This means that the next $s$ bytes indicate the length of the actual string which will then follow. This means that we have 8 bytes to encode the length of a string thereby giving us the maximum string length that can be encoded in RLP: $2 ^ {64} - 1$

\begin{lstlisting}[caption=RLP encoding of long strings, language=python, escapechar=\%, label={lst:rlp3}]
# We need one byte to represent 56
In [9]: rlp.encode('a' * 56)
Out[9]: b'\xb88aaaaaa...aa'

In [10]: 0xb8
Out[10]: 184

In [11]: # The 8 comes in because the ASCII
In [11]: # value of 8 is 56
In [11]: chr(56)
Out[11]: '8'

In [12]: rlp.encode('a' * 100)
Out[12]: b'\xb8daaaaaa...aa'

In [13]: chr(100)
Out[13]: 'd'

In [14]: rlp.encode('a' * 1000)
Out[14]: b'\xb9\x03\xe8aaaa...aaa'

In [15]: 0xb9 - 0xb7
Out[15]: 2

In [16]: 0x03e8
Out[16]: 1000
\end{lstlisting}

\item \textbf{192 to 247 ($0xc0$ to $0xf7$)}: The value now indicates that the combined length of the RLP encoding of each item of a list is between 0 and 55 bytes. The actual length is obtained by subtracting 192. This byte is followed by the RLP encoding of the individual items themselves which can be recursively decoded.

\begin{lstlisting}[caption=RLP encoding of a single byte, language=python, escapechar=\%, label={lst:rlp1}]
In [17]: rlp.encode(['a'])
Out[17]: b'\xc1a'

In [18]: 0xc1
Out[18]: 193

In [19]: rlp.encode(['a', 'b'])
Out[19]: b'\xc2ab'

In [20]: rlp.encode(['a', 'ab'])
Out[20]: b'\xc4a\x82ab'

In [21]: # Note the length is 4 bytes (0xc4)
In [21]: # because of the length byte
In [21]: # included from the RLP encoding of 'ab'
\end{lstlisting}

\item \textbf{248 to 255 ($0xf8$ to $0xff$)}: Similar to strings, these 8 values indicate that the next 1-8 bytes indicate the combined length of the RLP encodings of a list's individual items.

\begin{lstlisting}[caption=RLP encoding of a single byte, language=python, escapechar=\%, label={lst:rlp1}]
In [21]: rlp.encode(['a' * 56])
Out[21]: b'\xf8:\xb88aaaaaa...aa'

In [22]: len(rlp.encode('a' * 56))
Out[22]: 58

In [23]: chr(58)
Out[23]: ':'
\end{lstlisting}
\end{enumerate}

\chapter{Tries in Ethereum} \label{appendix:trie}

\section{Trie vs Tree}

\begin{figure}
\caption{A normal trie}
\label{img:trie}
\includegraphics[scale=0.6]{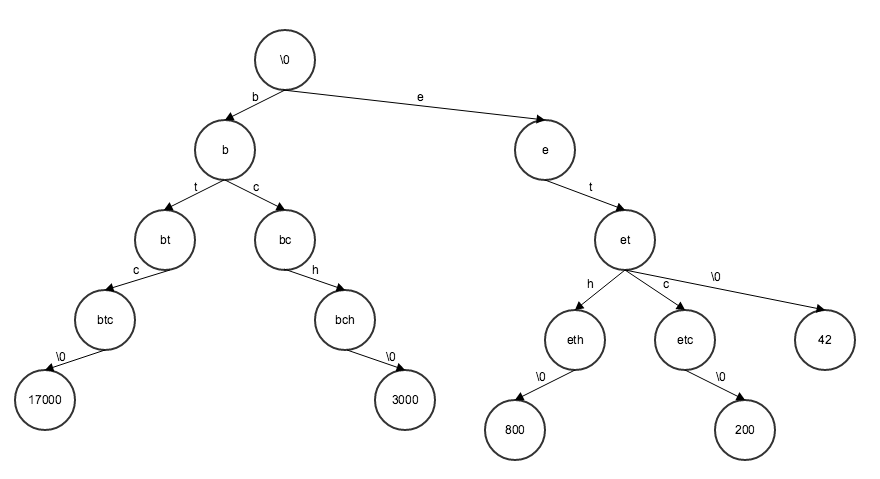}
\centering
\end{figure}

A trie is used to store key-value pairs among other things. We start with an empty key '' at the root of the trie. Traversing down a node increases the key by 1 character. A key can have a special child which does not append to the key but is rather the value of this key in the dictionary (the null child). A trie (Figure ~\ref{img:trie} is also known as a prefix tree because keys with the same prefixes (first k characters matching) are in the same sub-tree rooted at a node whose value is $s_1s_2...s_k$.

\section{Patricia Trie}
A Patricia trie optimizes a trie further by merging a child node with its parent if and only if the parent has no other children. When this happens, we can traverse multiple characters at one skip down a level instead of one character per level in a regular trie.

\begin{figure}
\caption{Patricia trie}
\label{img:patricia_trie}
\includegraphics[scale=0.6]{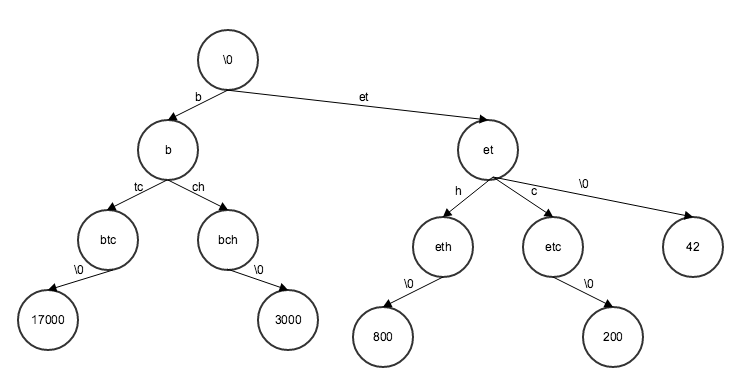}
\centering
\end{figure}

Figure ~\ref{img:patricia_trie} is the Patricia trie version of the trie in Figure ~\ref{img:trie}
\section{Merkle Patricia Trie}
In Ethereum, keys in the trie are 32-byte hashes. The Merkle aspect comes in because when a node refers to it child, it uses the keccak 256 hash of the RLP encoding of the value of its child. This reference is then included as a part of the parent's RLP encoding i.e the parent's parent includes the parent's hash in its value hence building a Merkle Tree.

Keys are represented using the hexadecimal format in the trie which means each key has 64 nibbles. Each traversal down a level usually adds one nibble to the key, unless we hit a special type of node which advances it by multiple nibbles (the Patricia part).

Before we delve into the different types of nodes, a note on the value stored in the leaves. When an object is to be inserted into the trie, for example a transaction, its RLP encoding is computed and then its keccak 256 (32 byte) hash. This (hash, RLP encoding of value) pair is stored in a key value datastore. This applies to a node in the trie itself. When a parent refers to a child, it is the address of the child in this datastore whose value is the RLP encoding of the child node itself (see example for leaf node below to see how this works)

Tries are used in Ethereum to maintain account related state (balances etc.), smart contract state, transactions and transaction receipts for each block.

\section{Types of nodes in an Ethereum Trie}
\begin{enumerate}
    
\item A NULL node represented by an empty string:
\begin{lstlisting}[caption=NULL node, basicstyle=\small, language=Python, escapechar=\#, label={lst:trie-null-node},showstringspaces=false]
In [1]: import trie

In [2]: t = trie.Trie(db={})

In [3]: t.root_node
Out[3]: b`'

In [4]: t.root_hash
Out[4]: b`V\xe8\x1f\x17...\xb4!'

In [5]: keccak(rlp.encode(b`'))
Out[5]: b`V\xe8\x1f\x17\x1b\xccU...\xb4!'

In [6]: t.db
Out[6]: {}
\end{lstlisting}

\item A leaf node represented by a two-item list: item 1 is the path traversed to get here i.e. the key and item 2 is the value.
\begin{lstlisting}[caption=NULL node, basicstyle=\small, language=Python, escapechar=\#, label={lst:trie-leaf-node}, showstringspaces=false]
In [7]: t = trie.Trie(db={})

In [8]: t.set(b`1', b'abc')

In [9]: t.root_node
Out[9]: [b` 1', b`abc']

In [10]: t.root_hash
Out[10]: b`\x1ft\xe9Y\xac\xcd&+...\xcaf'

In [11]: t.db
Out[11]: {b`\x1ft\xe9Y\...\xcaf': b`\xc7\x82 1\x83abc'}

In [12]: rlp.encode([b` 1', b`abc'])
Out[12]: b`\xc7\x82 1\x83abc'

In [13]: rlp.encode(b`abc')
Out[13]: b`\x83abc'

In [14]: keccak(rlp.encode([b` 1', b`abc']))
Out[14]: b`\x1ft\xe9Y\xac\xcd&+...\xcaf'
\end{lstlisting}

You'll notice that the path in the leaf node is ` 1' and not `1'. The space is inserted because a 'hex prefix' is added as the first nibble of the trie path. This prefix is used to differentiate between a leaf node and an extension node (covered next) and to also signal if the number of nibbles in the path is odd or even. If there are an even number of nibbles in the path, a `0' (0000) nibble is suffixed to the hex prefix nibble because we cannot have a dangling byte made up of only one nibble. The current pre-defined values for the hex prefix nibble are:
\begin{enumerate}
\item \textbf{3 (0011)} - Leaf node with odd number of nibbles (does not need a `0' nibble suffix between the hex prefix nibble evens things out)
\item \textbf{2 (0010)} - Leaf node with even number of nibbles (needs a `0' nibble suffix)
\item \textbf{1 (0001)} - Extension node with odd number of nibbles (does not need a `0' nibble suffix)
\item \textbf{0 (0000)} - Extension node with even number of nibbles (needs a `0' nibble suffix)
\end{enumerate}

With that out of the way, how did we then end up with ` 1'? The comments in the code sample below attempt an explanation:
\begin{lstlisting}[caption=NULL node, basicstyle=\small, language=Python, label={lst:trie-hex-prefix}, showstringspaces=false]
# The key `1' when converted to a nibble path becomes 
# the nibbles `3' and `1' because 31 is the
# hexadecimal equivalent of the ASCII value # of 1 (49)
In [15]: trie.utils.nibbles.bytes_to_nibbles(b`1')
Out[15]: (3, 1)

# Since we have an even number of nibbles (2) and 
# this is a leaf node, our hex prefix is 2 and we 
# add the 0 nibble suffix
In [15]: trie.utils.nibbles.nibbles_to_bytes((2, 0, 3, 1))
Out[15]: b` 1'

# 20 is the hexadecimal representation of the number
# 32 which is the ASCII value of whitespace ` '
In [16]: int(`20', base=16)
Out[16]: 32

In [17]: chr(32)
Out[17]: ` '
\end{lstlisting}

\end{enumerate}

\end{document}